\pdfoutput=1
\documentclass{article}

\PassOptionsToPackage{numbers,sort&compress}{natbib}
\usepackage[final,eandd]{template/neurips_2026}

\usepackage[utf8]{inputenc}
\usepackage[T1]{fontenc}
\usepackage{hyperref}
\hypersetup{
  colorlinks=true,
  linkcolor=purple,
  citecolor=blue,
  urlcolor=magenta,
}
\usepackage{url}
\usepackage{booktabs}
\usepackage{amsfonts}
\usepackage{nicefrac}
\usepackage{microtype}
\usepackage{xcolor}

\makeatletter
\newcommand{\authorfnmark}{\renewcommand{\@makefnmark}{\hbox{\normalfont\textsuperscript{\@thefnmark}}}}
\makeatother

\usepackage{epsfig,endnotes}
\usepackage{amsmath,amsthm}
\ifx\mystyle\usenix \usepackage{amssymb} \fi
\usepackage[ruled, noend, vlined]{algorithm2e}
\usepackage{listings}
\usepackage{xcolor}
\usepackage{colortbl}
\usepackage{balance}
\usepackage{graphicx}
\usepackage{subcaption}
\usepackage{wrapfig}
\usepackage{booktabs}
\usepackage[export]{adjustbox}
\usepackage{datetime}
\usepackage{enumitem}
\usepackage{multirow}
\usepackage{fvextra}
\usepackage{siunitx}
\usepackage{pifont}
\usepackage{tikz}
\usetikzlibrary{tikzmark,calc}
\usepackage{stackengine}
\usepackage[nameinlink]{cleveref}
\usepackage[title]{appendix}
\usepackage[breakable,skins]{tcolorbox}

\graphicspath{{./}{figure/}}

\definecolor{hlcolor}{RGB}{254,251,190}

\fvset{
  breaklines=true,
  breaksymbolleft=,
  fontsize=\footnotesize,
}

\newcommand{\bench}{\textsc{DistDebug-Bench}\xspace}
\newcommand{\benchrepo}{\url{https://github.com/OrderLab/distdebug-bench}\xspace}
\newcommand{\benchsite}{\url{https://distdebug.com}\xspace}
\newcommand{\tracesurl}{\url{https://huggingface.co/datasets/OrderLab/distdebug-bench-traces}\xspace}
\newcommand{\benchsize}{120\xspace}
\newcommand{\numsystems}{11\xspace}
\newcommand{\numlanguages}{5\xspace}
\newcommand{\numpatterns}{10\xspace}

\DeclareMathAlphabet{\mathcal}{OMS}{cmsy}{m}{n}

\newcommand{\findinghdr}[2]{\vspace{0.05in}\noindent{\textbf{#1}\enspace{\textit{#2}}}}

\definecolor{findingbg}{RGB}{237,245,253}
\definecolor{findingframe}{RGB}{30,19,255}

\newcounter{finding}
\newcommand{\findingbox}[2]{%
  \refstepcounter{finding}%
  \par\smallskip%
  \begin{tcolorbox}[
    enhanced,
    colback=findingbg,
    colframe=findingframe,
    boxrule=1pt,
    arc=2pt,
    left=0.6em, right=0.6em,
    top=0.45em, bottom=0.45em,
    breakable,
  ]%
  \textbf{Finding~\thefinding.}\enspace\emph{#1.} #2%
  \end{tcolorbox}%
  \par\smallskip%
}

\DeclareSIUnit{\operation}{op}
\definecolor{ngray}{RGB}{102,102,102}

\newcolumntype{R}[2]{%
  >{\adjustbox{angle=#1,lap=\width-(#2)}\bgroup}%
  l%
  <{\egroup}%
}

\definecolor{bg}{rgb}{0.95,0.95,0.95}

\crefname{appendix}{appendix}{appendices}
\Crefname{appendix}{Appendix}{Appendices}

\newcounter{magicrownumbers}

\title{\bench: Can LLM Agents Diagnose Bugs in Distributed Systems?}

\author{%
  \authorfnmark
  \begingroup
  \hypersetup{urlcolor=black}%
  \begin{tabular}{ccc}
    \href{https://essoz.github.io}{Yuxuan Jiang} &
    Chenglin Liu\thanks{Work done while at the University of Michigan.\\ Correspondence to: \texttt{jyuxuan@umich.edu}, \texttt{ryanph@umich.edu}.} &
    John Rose\footnotemark[1] \\[1.1ex]
    Mason Miller &
    Ruthesh Thavamani &
    Juntao Wu \\[1.1ex]
    Yinwei Huang\footnotemark[1] &
    Jiahao Xu &
    \href{https://web.eecs.umich.edu/~ryanph}{Peng Huang}
  \end{tabular}%
  \endgroup
  \\[6ex]
  {\normalfont\normalsize University of Michigan}\\[0.55ex]
}

\begin{document}

\maketitle

\begin{abstract}
  LLM agents show strong potential in software engineering tasks such as
  competitive programming and code synthesis. Existing benchmarks, however,
  largely focus on small-scale settings such as self-contained Python modules,
  leaving distributed systems, an essential class of production software,
  under-examined. We introduce \bench, a benchmark for evaluating agents' ability to
  diagnose complex-system failures at the code level. It curates
  \benchsize high-quality real-world bugs from \numsystems widely used
  distributed systems, spanning \numlanguages programming languages and
  \numpatterns failure patterns. Through an extensive evaluation of state-of-the-art
  LLMs and agents, we identify substantial gaps between current agents and
  real-world debugging tasks. A recurring gap is tunnel vision due to
  a lack of explicit reasoning structure and nondeterminism from
  inconsistent tool-use scope. The benchmark and evaluator are available at
  \benchrepo, with agent traces and a leaderboard at \benchsite.
\end{abstract}

\section{Introduction}\label{sec:intro}
Large language models (LLMs) and agents have made rapid progress on software
engineering tasks, including code synthesis, repository navigation, and
issue resolution. Early code benchmarks measure LLM capabilities on
short-program synthesis~\cite{chen2021codex,austin2021programsynthesislargelanguage};
programming-contest benchmarks evaluate more challenging algorithmic problem
solving~\cite{jain2024livecodebenchholisticcontaminationfree,zheng2025livecodebenchproolympiadmedalists};
RepoBench measures cross-file repository-context use
\citep{liu2023repobenchbenchmarkingrepositorylevelcode}; later benchmarks
such as SWE-bench and SWE-bench Verified evaluate whether an agent can resolve
GitHub issues by producing a patch that passes a test
suite~\cite{jimenez2024swebench,chowdhury2024swebenchverified}.
These benchmarks help shape the designs of both models and agent harnesses.

However, much existing benchmarking focuses on small-scale or locally scoped
settings, such as standalone functions, self-contained modules, and
library-style repositories. These settings are useful, but leave a critical
real-world target under-examined: \emph{debugging large-scale, production
systems}. Existing evaluations also often emphasize patch generation, which can
conflate diagnosis with repair synthesis. A patch may suppress a symptom
without addressing the root cause; even when the final fix changes only a few
lines, finding that cause may require understanding a much larger causal chain.
In many production workflows, engineers must therefore diagnose a failure
before writing a proper fix. This debugging workflow is crucial yet not
directly measured by existing benchmarks.

\emph{Distributed systems debugging} is a particularly important and
challenging setting in this gap. Production distributed systems, including
distributed storage, coordination, and stream processing, underpin modern
computing services. Their failures are costly and hard to debug for several
reasons. First, these systems span millions of lines of code across many subsystems and
layered abstractions, creating a large search space in which the true root
cause occupies only a tiny subset. Second, failures involve interactions across
modules, network calls, and service tiers. An observed symptom often exposes
only downstream effects while the originating fault is far away. Third, the
available evidence is noisy and incomplete: issue reports, logs, stack traces,
and artifacts provide only \emph{partial observations}. Large-scale
distributed systems also experience gray failures with ambiguous
symptoms~\cite{HuangHotOS17,watchdog2020nsdi}. Debugging therefore requires
reasoning from high-level symptoms to low-level code under weak evidence.
Developers need to generate and test competing hypotheses, reconcile evidence with
code context, trace causal relationships across modules and historical design
decisions, and revise assumptions as new evidence appears. Distributed-system
debugging thus stresses capabilities central to LLM and agent design:
\emph{long-context retrieval}, \emph{large-repository navigation},
\emph{multi-hop causal reasoning}, \emph{decision-making under noisy and
incomplete evidence}, and \emph{belief revision}.

To address this gap, we introduce \textsc{DistDebug-Bench}, a benchmark for
evaluating LLM agents on \emph{code-level diagnosis of real-world
distributed-system failures}. In each benchmark instance, the agent is given an
incident report, accompanying evidence (\emph{e.g.}, stack trace), and the
pre-fix repository snapshot, and must identify one or more source-code blocks
that best explain the observed failure. Outputs are judged against
human-annotated labels that capture the root-cause regions.
The benchmark contains \benchsize curated incidents from \numsystems
distributed systems and \numlanguages languages, with 219 gold root-cause
blocks. It captures the properties that make distributed-system debugging
difficult: 46\% require multiple blocks to express the root cause, 24\% span
multiple files, 68\% of reports do not mention the gold method, and 63\%
contain no usable internal stack trace. The benchmark repository is available
at \benchrepo, with project documentation and updates at \benchsite.

We evaluate state-of-the-art LLMs and agents on the benchmark.
The results reveal a substantial gap between broad repository navigation and
precise diagnosis. The strongest configuration achieves high file-level
coverage but much lower block-level coverage and precision. Agents often find a
relevant component but fail to pinpoint the causal source lines. Performance
degrades most for opaque reports and cross-component bugs. Through trace inspection,
we find a recurring failure mode: agents commit early to a plausible local
explanation and rarely recover when the initial search scope misses the causal
region. We call this behavior \emph{tunnel vision} and analyze how it arises
from evidence structure and tool-call trajectories. Non-determinism further
compounds the problem: even a leading solution produces flaky
diagnoses on 19\% of incidents, rising to 29\% on no-stack opaque reports.

Our contributions are:
\begin{itemize}[itemsep=0pt, topsep=1pt, partopsep=0pt, leftmargin=*]
    \item A benchmark task for distributed-system diagnosis in which agents
      localize the pre-fix code blocks that best explain an observed failure.
    \item A curated dataset of real-world failures in widely used distributed
      systems, with a label design process focused on reproducibility and
      non-trivial causal structure.
    \item An empirical evaluation of current LLM agents and harnesses showing
      gaps in precise causal localization, especially for weak-evidence and
      cross-component failures.
    \item A failure-mode analysis identifying key obstacles for reliable
      debugging.
\end{itemize}

\section{Related Work}
\label{sec:related}

\textbf{Code-generation and repository-context benchmarks.}
Early code benchmarks such as HumanEval and MBPP evaluate functional
correctness for short programming tasks~\cite{chen2021codex,
austin2021programsynthesislargelanguage}. LiveCodeBench and LiveCodeBench Pro
extend this line with newer competitive-programming tasks and broader
code-execution signals~\cite{jain2024livecodebenchholisticcontaminationfree,
zheng2025livecodebenchproolympiadmedalists}. Repository-context benchmarks such
as RepoBench move beyond single-file synthesis by testing retrieval and
completion across files~\cite{liu2023repobenchbenchmarkingrepositorylevelcode}.
These benchmarks measure code synthesis abilities but not failure diagnosis.

\textbf{Bug, repair, and debugging datasets.}
Long-running software-engineering datasets such as Defects4J, ManyBugs,
Bugs.jar, Bears, BugSwarm, and BugsInPy provide reproducible real bugs and
failing and passing versions for testing, fault localization, and automated program
repair~\cite{just2014defects4j,legoues2015manybugsintroclass,saha2018bugsjar,
madeiral2019bears,tomassi2019bugswarmminingcontinuouslygrowing,
widyasari2024bugsinpydatabaseexistingbugs}. More recent LLM debugging
benchmarks such as DebugBench directly test whether models can repair or reason
about \emph{injected} buggy snippets~\cite{tian2024debugbenchevaluatingdebuggingcapability}.
In these datasets, the primary oracle is typically a test suite, a fixed
version, or a snippet-level repair target; diagnosis is not directly evaluated.
\bench explicitly measures root-cause localization. Agents must identify the
precise causal block(s). It targets large-scale distributed systems. Its bugs
are drawn from challenging, real-world failures. Its diagnosis
tasks model the multi-source, noisy evidence of a real incident.

\textbf{Issue-resolution agent benchmarks.}
SWE-bench changed the evaluation target from standalone synthesis to real
GitHub issue resolution, and follow-on suites add human verification, visual
software domains, live task collection, multilingual ecosystems, broader
repository languages, and long-horizon enterprise-style tasks~\cite{
jimenez2024swebench,chowdhury2024swebenchverified,
yang2024swebenchmultimodalaisystems,zhang2025swebenchgoeslive,
zan2025multiswebenchmultilingualbenchmarkissue,
rashid2025swepolybenchmultilanguagebenchmarkrepository,
deng2025swebenchproaiagents}. SWE-Gym further targets training environments and
verifiers for software-engineering agents~\cite{pan2025trainingsoftwareengineeringagents}.
\bench shares the basic format (repository-level context and natural-language
issue reports) but differs along several axes: (1) they score patches, while
\bench scores root-cause localization directly; (2) SWE-bench-style instances
are dominated by single-process projects with short causal chains; (3) they
require a test suite for evaluation, a constraint that biases curation toward bugs whose
fixes are exercised by existing tests. For many distributed systems failures, a
regression test rarely exists at the time of diagnosis.

\textbf{Operational benchmarks.} Recent benchmarks evaluate LLM
agents on IT automation and fault localization. AIOpsLab evaluates AIOps
agents in microservices with fault injection, workload generation, and telemetry
export~\cite{chen2025aiopslabholisticframeworkevaluate}.
ITBench evaluates AI agents across IT automation tasks~\cite{jha2025itbenchevaluatingaiagents}.
SREGym evaluates SRE agents in a controlled platform~\cite{sregym}.
Benchmarks such as OpenRCA~\cite{xu2025openrca}, RCAEval~\cite{pham2025rcaevalbenchmarkrootcause},
and PetShop~\cite{hardt2024petshopdatasetfinding} evaluate LLMs on fault localization: identifying the faulty
services, machines, or resources from telemetry, metrics, traces, and injected
performance anomalies. These benchmarks are valuable for
AIOps~\cite{AIOps2019ICSE} workflows. They typically operate over telemetry
data and focus on coarse-grained localization. \bench addresses a complementary
problem: \emph{precise source-level debugging}. In incident response,
engineers still need to perform the
last-mile reasoning that connects symptoms to the root cause block(s) in code,
which is often the most time-consuming step. \bench evaluates LLMs'
abilities in this task. It preserves repository-level, source-code
grounding while capturing the weak evidence, interactions, and ambiguity
characteristic of distributed-system debugging.

\section{Benchmark}
\label{sec:benchmark}

\begin{figure}[t]
  \centering
  \includegraphics[width=\textwidth]{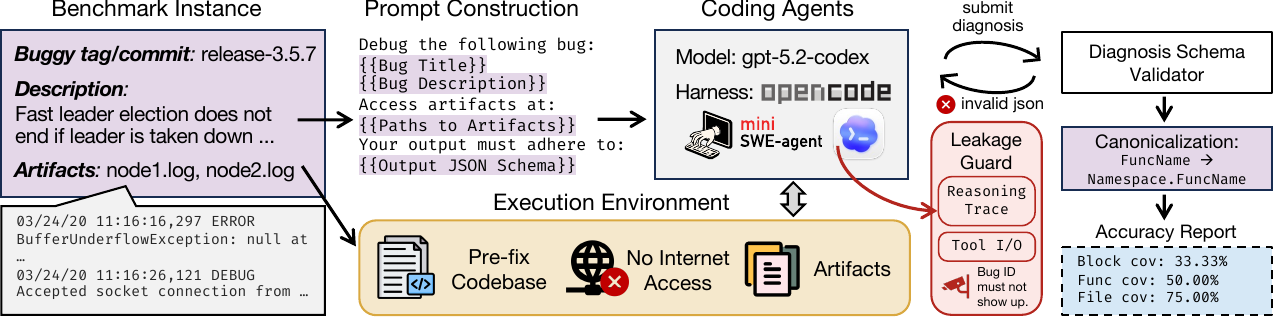}
  \caption{\bench task flow.} %
  \label{fig:teaser}
\end{figure}

\bench is a benchmark for \emph{code-level diagnosis} of real-world
distributed-system failures.
\bench evaluates coding agents in the real-world systems maintenance workflow
where developers have to start with a high-level symptom, explore possible root
causes by forming and testing multiple causal hypotheses from partial evidence,
and identify where in the code the root cause is encoded before any fix is applied.
Each benchmark instance asks the agent to identify, from an incident report, a realistic
evidence bundle, and the pre-fix snapshot, the code block or set of blocks that
best explains the observed failure.

\subsection{Task Workflow}

\Cref{fig:teaser} shows the task flow. The agent-facing input consists of an
incident report, issue-side artifacts such as logs or stack traces, the pre-fix
repository snapshot, and a structured output schema. The fixing change and
human root-cause labels are held out for annotation and scoring. The agent must
return one or more repository-local source blocks that encode the pre-fix
failure mechanism. The evaluator validates the structured output, canonicalizes
symbols, and scores whether the predicted blocks capture the minimal causal
region(s) that explain the incident.

The benchmark release contains \benchsize incidents from \numsystems
production-scale distributed systems across \numlanguages languages. Each
incident is selected to be diagnosable from issue-side context, stripped of
answer-leaking discussion, and labeled with precise pre-fix root-cause blocks.
Because \bench is built from public historical bugs, we also include leakage
controls in the task workflow. The prepared workspaces anonymize directory names
and remove commits after the task snapshot. The leakage guard in
\Cref{fig:teaser} scans visible reasoning and tool inputs/outputs for the original
issue identifiers. Flagged attempts are discarded and rerun;
\Cref{app:benchmark-validity} describes the guard and follow-up audit process.
For GPT-5.4, flagged attempts fell from 102/360 to 4/360 after future-commit
removal. In closed-book probes on 66 reports without exact gold file/function
names, revealing the issue ID gives no advantage: function coverage is 17.1\%
with the ID and 21.1\% without it, versus 46.7\% with repository access.
Repository access substantially
improves localization in these probes, although pretraining exposure cannot be ruled out
(\Cref{app:contamination-evidence}).

\subsection{Bug Selection}

\bench is constructed from issue reports and fixing changes drawn from
large-scale, production distributed systems. We select candidate bugs using
three principles: the issue should describe a representative production failure,
the available evidence should support diagnosis without revealing the answer,
and the fixing change should support precise root-cause labeling.

\textbf{Real production failures.} We prioritize high-impact, user-facing bugs
on stable releases. Eligible issues must be closed, have an associated fix, and
describe a concrete failure rather than a feature request or internal cleanup.

\textbf{Realistic and diagnosable evidence.} We prefer reports that expose the
kinds of signals engineers actually use in practice: stack traces, logs,
reproduction steps, cross-linked prior issues, or explicit symptom descriptions.
We also check that the issue is diagnosable from issue-side context alone. Later
issue-thread comments that reveal the root cause or patch strategy are removed
or ignored, since those no longer reflect realistic diagnosis.

\textbf{Labelable root-cause fixes.} We retain bugs whose fixes are local enough
to support precise root-cause labeling. Large refactorings, test-only changes,
and patches dominated by unrelated cleanup are excluded because they make the
minimal causal span ambiguous. As a practical filter, we also exclude fixes with
more than 100 changed source lines unless the causal span remains clearly
separable from auxiliary edits.

\subsection{Root-Cause Labeling}

For each accepted issue, annotators derive the minimal pre-patch root-cause span
from the fixing change and its surrounding context. Each label marks a buggy
code block with a file path, function or method symbol, and a start/end line
range in the pre-fix snapshot. \Cref{app:benchmark-artifacts} shows the
concrete label format.

In practice, annotators inspect the issue report, the fixing patch, and the
affected code, and then mark only the code that embodied the bug before the fix.
We exclude tests, helper refactors, defensive hardening, and incidental cleanup
unless these are necessary parts of the minimal causal explanation, because such
edits do not directly encode the original failure mechanism. Each issue is
independently annotated by three annotators and then reconciled into a final
gold label set; \Cref{app:annotation-agreement} reports inter-annotator
agreement.

In addition to root-cause labels, we record characterization metadata for each
bug. These fields include textual code anchors, crash-to-fix distance,
issue-side signals, and coarse root-cause type. They are not scoring targets or
agent-facing hints, but they provide the foundation for the fine-grained
breakdowns in \Cref{sec:evaluation}.

\textbf{Scaling collection.} The released \bench labels are fully human-curated
and human-reconciled; LLM outputs are not used in the gold annotation workflow.
Because human review remains the main bottleneck for scaling, we also developed
an LLM-assisted label extraction pipeline for future exploration.
\Cref{app:collection-pipeline-eval} describes it and reports a small-subset
comparison between this pipeline, a non-LLM extraction pipeline, and the final
human labels.

\subsection{Benchmark Characteristics}
\label{sec:bench-characteristics}

\begin{figure}[t]
  \centering
  \includegraphics[width=\textwidth]{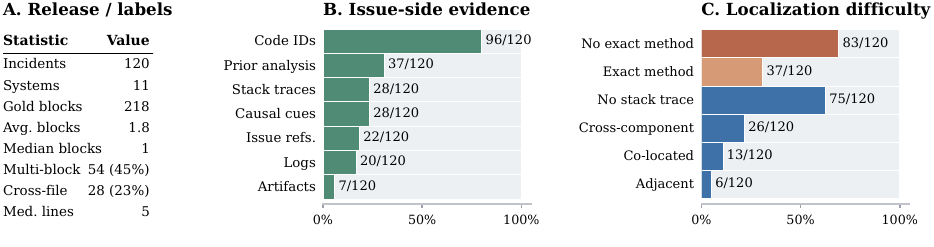}
  \caption{\bench workload characterization.} %
  \label{fig:bench-characterization}
\end{figure}

\Cref{fig:bench-characterization} summarizes the workload beyond headline
dataset size. The characterization highlights three properties that make
diagnosis difficult and later enable fine-grained evaluation breakdowns: the
issue-side evidence is heterogeneous, the symptom is often far from the causal
code, and many gold labels require non-local or multi-block explanations.

\textbf{Issue-side evidence and anchors.}
\bench reports contain heterogeneous evidence rather than a uniform problem
statement. Code identifiers appear in 96 / 120 reports, but direct debugging
signals are much less common: 37 include prior analysis, 28 include stack
traces, 20 include logs, and only 7 include attached artifacts. Even when a
report contains concrete code names, it rarely names the final fix location: in
82 / 120 reports (68.3\%), the exact gold method does not appear in the title or
description.

\textbf{Crash-to-fix distance.}
Stack traces are also an incomplete localization signal. In 75 / 120 incidents,
the report does not contain a usable project-internal stack trace. Among the
remaining incidents with an observable crash site, only 13 are co-located with
the fix, while 26 require crossing component boundaries from the crash site to
the gold block and 6 are adjacent within nearby package structure.

\textbf{Ground-truth structure.}
The labels also show that diagnosis is not simply a single-location search
problem. \bench contains 219 gold buggy blocks, and 45.8\% of incidents require
more than one block to express the root cause. Nearly a quarter span multiple
files. At the same time, the median block is only four lines, so a useful agent
must both follow non-local causal chains and report precise source spans.
\Cref{app:benchmark-details} gives the full artifact layout, raw instance
examples, label schema, and classification rules used for the benchmark
characterization.
For a concrete illustration of how issue evidence can point to a plausible
symptom-side investigation space while the actual root cause lies elsewhere,
\Cref{app:zk2845-example} walks through ZooKeeper-2845.

\section{Evaluation}\label{sec:evaluation}

We evaluate whether current software-engineering agents can identify
source-level root causes for \bench incidents, and where their diagnosis quality
breaks down. The central result is that contemporary agents often reach relevant
code, but their diagnosis process breaks down when weak incident evidence
requires sustained navigation and hypothesis revision.

\subsection{Evaluation Setup}

We run \bench along two axes: (1) the model axis and (2) the harness axis. We
first conduct a thirteen-configuration sweep on Mini-SWE-Agent, including
GPT-6 Astra, Claude Opus 5.5, GPT-5.5, GPT-5.4, GPT-5.3-Codex, GPT-5.2-Codex, Hunyuan3, o4-mini,
DeepSeek V4 Flash, Kimi K2.6, Gemini 3.5 Flash, MiMo v2.5, and MiMo v2.5 Pro. We then run
GPT-5.2-Codex and o4-mini on Codex CLI and OpenCode to provide
harness comparison points. Mini-SWE-Agent is the common minimal harness used by
existing coding benchmarks~\citep{swebenchVerifiedLeaderboard,miniSweAgentDocs};
Codex CLI is OpenAI's terminal coding agent~\citep{openaiCodexCliDocs}; and
OpenCode is a highly starred open-source terminal coding agent~\citep{opencodeGithub}.
The model set mixes recent high-end systems, including GPT-6 Astra~\citep{openaiGpt6AstraDocs},
Claude Opus 5.5~\citep{anthropicClaudeOpus55Docs}, and
GPT-5.4~\citep{openaiGpt54Docs}, with popular economical or coding-specialized
options such as o4-mini, Kimi K2.6, DeepSeek V4 Flash, Gemini 3.5 Flash, and MiMo.

In addition to the 17 main configurations, we run exploratory experiments with
DeepSeek V4 Pro, Qwen3.6 Max, and Qwen3-Coder-Next on
non-uniform subsets of the benchmark. These runs are not included in the main
120-bug leaderboard, but \Cref{app:exploratory-subset-experiments} reports paired
comparisons against Mini-SWE GPT-5.2-Codex on the exact subset completed by each
model.

We set reasoning effort to medium when the model exposes that control. The MiMo
endpoint did not expose a reasoning-effort parameter, so those runs use the
provider default. Claude Opus 5.5 runs at its API default effort, which is
medium, with a 16,000-token output limit and Anthropic prompt caching. For
each bug/configuration pair, we average three attempts within the bug, then
average equally across all 120 bugs (\Cref{app:eval-attempt-counts}).
\Cref{app:eval-uncertainty} reports 95\% bug-level bootstrap confidence
intervals (CIs) for every configuration and paired comparisons between them.

\paragraph{Metrics.}
We compare generated diagnoses with labeled ground truth using coverage and
precision at four granularities: file, function, line, and block. File and
function metrics measure whether an agent reaches the right component. Line
coverage is the stricter overlap metric, while block coverage is a coarser
region-hit metric that asks whether each labeled causal block is reached. We use
block coverage as the main localization metric because it is less sensitive to
exact line-boundary choices and better captures whether an agent navigates to the
causal region. We report line coverage as a stricter sensitivity metric
(\Cref{app:metric-sensitivity}). We
also report \emph{mention coverage}: whether the natural-language
causal-chain explanation mentions gold functions even when the structured block
prediction is wrong. To measure nondeterminism, we compute the within-bug
standard deviation of block coverage across the recorded attempts and the number
of \emph{flaky} bugs whose block-coverage spread exceeds 0.2.

\paragraph{Metric validity.}
Block coverage measures recovery of the annotated causal code, not every
plausible intervention or repair point. Most zero-coverage diagnoses reflect
incorrect or incomplete causal reasoning: in a score-blinded audit of 20
zero-coverage GPT-5.4 diagnoses, 16 were incorrect or incomplete, and the other
four reconstructed the failure mechanism but blamed a different intervention
point. Benchmark-wide, most missed gold blocks are genuine misses: in a
stratified random sample of 92 incompletely covered attempts from all 17
configurations, two independent AI reviewers, each given the historical fix,
agree that 4.5\% (95\% CI 1.6--8.3\%) of the missed blocks were correctly
localized; counting them would raise block coverage by 2.3 points.
Coverage also matters downstream. In a six-bug within-bug repair pilot,
high-coverage diagnoses led the repair agent to edit the gold block in 50.0\% of
attempts, versus 11.1\% without a diagnosis and 0\% with low-coverage diagnoses
(\Cref{app:evaluator-validation}).

\begin{table*}[t]
  \centering
  \scriptsize
  \definecolor{evalmaincol}{RGB}{239,246,255}
  \definecolor{evaltoprow}{RGB}{248,251,255}
  \definecolor{evalgroupbg}{RGB}{238,241,245}
  \setlength{\tabcolsep}{3.4pt}
  \renewcommand{\arraystretch}{1.06}
  \caption{Main evaluation results across the 120 common \bench incidents. The shaded column is the primary localization metric, with 95\% bug-bootstrap CIs in brackets; bold values are column bests. \Cref{tab:eval-confidence-intervals} gives CIs for the other metrics.}
  \label{tab:eval}
  \resizebox{\textwidth}{!}{%
  \begin{tabular}{@{}ll@{\hspace{6pt}}rrr>{\columncolor{evalmaincol}}r@{\hspace{6pt}}rr@{\hspace{6pt}}r@{\hspace{6pt}}rrr@{}}
    \toprule
    \multicolumn{2}{@{}l}{\textbf{Configuration}} & \multicolumn{4}{c}{\textbf{Localization coverage}} & \multicolumn{2}{c}{\textbf{Explanation}} & \multicolumn{1}{c}{\textbf{Instability}} & \multicolumn{3}{c@{}}{\textbf{Effort}} \\
    \cmidrule(lr){1-2}\cmidrule(lr){3-6}\cmidrule(lr){7-8}\cmidrule(lr){9-9}\cmidrule(l){10-12}
    Model & Harness & File & Func. & Line & Block & Block prec. & Mention & Flaky bugs & Blocks/diag. & Calls/run & Tokens/run \\
    \midrule
    \rowcolor{evalgroupbg}\multicolumn{12}{@{}l@{}}{\textit{Mini-SWE-Agent model sweep}} \\
    \rowcolor{evaltoprow} GPT-6 Astra & Mini-SWE-Agent & 86.1 & 59.8 & 58.2 & \textbf{72.6}\,{\tiny[65.7, 79.3]} & 46.0 & \textbf{83.3} & \textbf{14 (12\%)} & 3.3 & 9.0 & \textbf{130K} \\
    Opus 5.5 & Mini-SWE-Agent & 84.5 & \textbf{61.8} & \textbf{58.9} & 71.8\,{\tiny[65.2, 78.2]} & 49.0 & 82.1 & 19 (16\%) & 2.8 & \textbf{8.8} & 149K \\
    GPT-5.5 & Mini-SWE-Agent & \textbf{86.2} & 57.4 & 57.7 & 70.0\,{\tiny[63.1, 76.7]} & 37.3 & 76.0 & 23 (19\%) & 3.6 & 22.0 & 474K \\
    GPT-5.4 & Mini-SWE-Agent & 81.0 & 54.6 & 53.6 & 66.6\,{\tiny[59.7, 73.4]} & 41.6 & 70.4 & 23 (19\%) & 2.9 & 25.0 & 687K \\
    5.3-Codex & Mini-SWE-Agent & 79.9 & 54.1 & 48.1 & 63.2\,{\tiny[55.9, 70.4]} & 42.7 & 69.9 & 25 (21\%) & 2.7 & 22.7 & 417K \\
    MiMo v2.5 Pro & Mini-SWE-Agent & 80.5 & 60.6 & 46.7 & 62.0\,{\tiny[54.5, 69.3]} & 40.2 & 73.5 & 22 (18\%) & 2.9 & 55.1 & 2617K \\
    Kimi K2.6 & Mini-SWE-Agent & 79.2 & 51.9 & 48.0 & 62.0\,{\tiny[54.8, 69.0]} & 49.2 & 71.4 & 33 (28\%) & 2.4 & 66.0 & 2784K \\
    MiMo v2.5 & Mini-SWE-Agent & 79.7 & 54.9 & 46.9 & 61.1\,{\tiny[53.8, 68.2]} & 38.7 & 70.4 & 27 (22\%) & 3.1 & 48.7 & 2265K \\
    5.2-Codex & Mini-SWE-Agent & 76.7 & 49.6 & 49.2 & 59.5\,{\tiny[51.7, 67.1]} & 45.5 & 64.4 & 19 (16\%) & 2.3 & 31.9 & 665K \\
    DS V4 Flash & Mini-SWE-Agent & 77.4 & 50.1 & 48.6 & 59.2\,{\tiny[52.0, 66.2]} & 34.6 & 68.7 & 36 (30\%) & 3.2 & 62.1 & 2323K \\
    HY3 & Mini-SWE-Agent & 74.1 & 49.8 & 42.5 & 54.8\,{\tiny[47.4, 62.2]} & 41.3 & 65.3 & 36 (30\%) & 2.7 & 32.3 & 761K \\
    Gemini 3.5 & Mini-SWE-Agent & 67.7 & 44.4 & 40.7 & 50.8\,{\tiny[43.1, 58.3]} & \textbf{54.5} & 59.7 & 26 (22\%) & 1.4 & 25.2 & 520K \\
    o4-mini & Mini-SWE-Agent & 64.2 & 36.3 & 31.3 & 42.1\,{\tiny[34.9, 49.6]} & 37.3 & 50.7 & 33 (28\%) & 1.8 & 29.9 & 498K \\
    \addlinespace[0.25ex]
    \rowcolor{evalgroupbg}\multicolumn{12}{@{}l@{}}{\textit{Harness comparison}} \\
    5.2-Codex & Codex CLI & 65.7 & 47.7 & 38.7 & 49.9\,{\tiny[42.0, 57.7]} & 52.7 & 60.2 & 19 (16\%) & 1.6 & N/A & 494K \\
    o4-mini & Codex CLI & 56.5 & 37.5 & 25.3 & 35.9\,{\tiny[28.6, 43.5]} & 35.5 & 52.0 & 28 (23\%) & 1.7 & N/A & 331K \\
    5.2-Codex & OpenCode & 69.4 & 45.3 & 39.0 & 51.4\,{\tiny[43.7, 59.1]} & 49.3 & 58.6 & 22 (18\%) & 1.8 & 18.4 & 547K \\
    o4-mini & OpenCode & 55.0 & 31.3 & 24.8 & 35.1\,{\tiny[28.0, 42.2]} & 40.5 & 45.4 & 27 (22\%) & \textbf{1.4} & 13.5 & 280K \\
    \bottomrule
  \end{tabular}
  }
  \vspace{0.02in}
  \begin{minipage}{0.98\textwidth}
  \footnotesize\emph{Note.} Values are averaged over 120 common bugs, after averaging the first three valid scored attempts within each bug. Coverage, block precision, mention coverage, and flaky-bug rates are percentages. \Cref{app:eval-attempt-counts} details attempt selection and the common cohort. Lower is better for flaky bugs, blocks/diagnosis, calls/run, and tokens/run. Flaky bugs have block-coverage spread above 0.2 across attempts. 5.3-Codex and 5.2-Codex abbreviate GPT-5.3-Codex and GPT-5.2-Codex.
  \end{minipage}
\end{table*}

\subsection{Main Evaluation Results}

\Cref{tab:eval} summarizes diagnosis performance across the main model and
harness configurations. GPT-6 Astra has the highest aggregate block coverage in our complete sweep: with
Mini-SWE, it reaches 86.1\% file coverage, 58.2\% line coverage, and 72.6\%
block coverage. Across configurations, however, file coverage remains
substantially higher than line and block coverage, and higher-coverage settings
often come with broader predictions: Mini-SWE GPT-5.5 has 37.3\% block
precision and names 3.6 blocks per diagnosis on average. These aggregate results
point to a consistent theme:
\textbf{Current SOTA agents can often reach relevant code, but their diagnosis
process breaks down when weak incident evidence requires sustained navigation
and hypothesis revision.} The category breakdown below makes this gap visible:
the weakest cases are not only those without starting evidence, but also those
where the report provides an initial symptom location without naming the causal
code.

\begin{figure}[t]
  \centering
  \includegraphics[width=\textwidth]{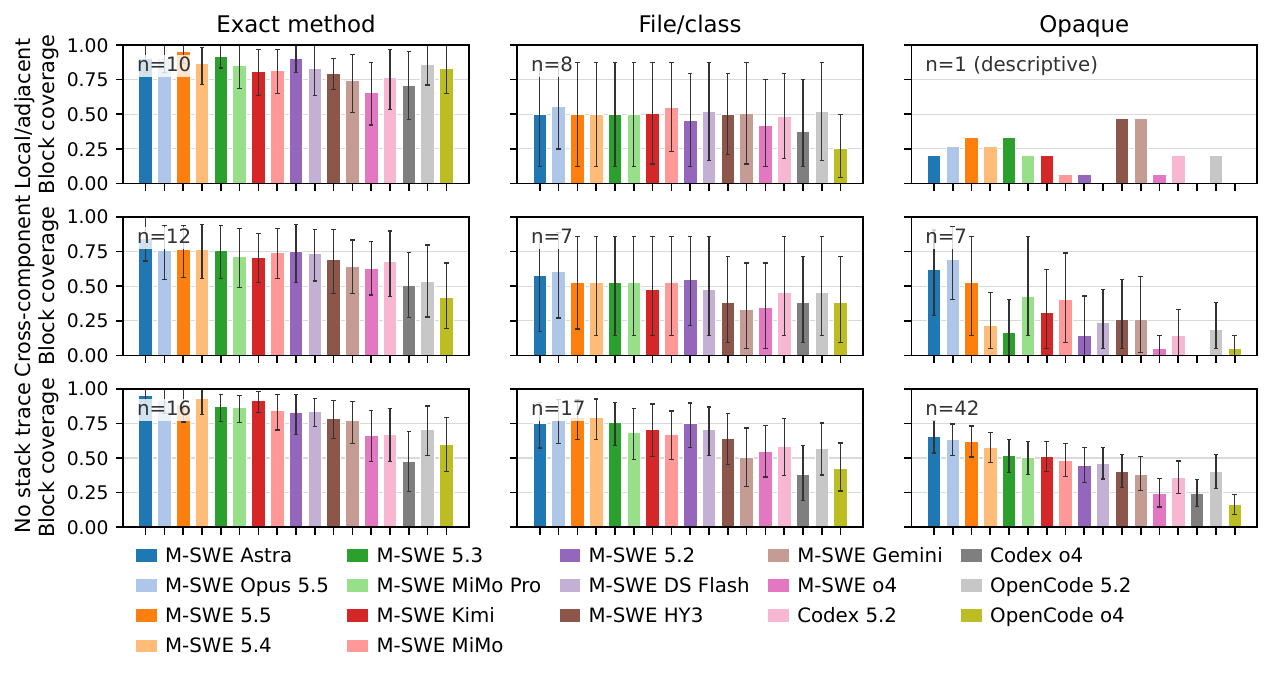}
  \caption{Block coverage by incident evidence. Columns group the issue text's
  strongest code anchor; rows group the first crash stack frame. Coverage drops
most when reports lack a precise anchor and the crash site is absent or
cross-component. Panel counts
give bugs per cell; whiskers show 95\% bug-bootstrap CIs (none for the
single-bug cell).}
  \label{fig:eval-evidence-block-coverage}
\end{figure}

\findingbox{Agents struggle when the report does not point to the exact code
locations}{Performance drops sharply when the bug report does not mention the
ground-truth function names or files, suggesting that the report's initial
content heavily shapes the final diagnosis and that agents have limited ability
to revise early hypotheses.}
\Cref{app:failure-mode-details} gives a concrete trace example for this
failure mode.

\Cref{fig:eval-evidence-block-coverage} divides the bugs into 9
categories, based on two kinds of help that an incident report can give the
agent: whether the text names an exact gold method, a gold file/class, or no
causal code at all; and whether the first crash stack frame is local or adjacent
to the fix, cross-component, or absent. The figure reports per-cell sample
counts and 95\% bug-bootstrap CIs; small-cell comparisons are exploratory.

When the report names an exact method, agents often do well: the best current
configuration reaches 95.0\% block coverage on local-or-adjacent method-named
bugs, and 94.8\% on no-stack method-named bugs. When the report is opaque, even
the best configuration reaches only 65.6\% on the 42 no-stack opaque bugs, and
12 of the 17 configurations stay below 40\% on the seven cross-component opaque bugs.
For a fixed baseline, Mini-SWE GPT-5.2-Codex shows the same pattern: its
no-stack method-named coverage is 82.8\%, but it falls to 44.9\% on no-stack
opaque bugs and 14.3\% on cross-component opaque bugs.

Adding newer models raises absolute coverage but does not remove this weak-evidence
gap. GPT-6 Astra improves no-stack opaque block coverage to 65.6\%, yet this remains
29.2 percentage points below its 94.8\% coverage on no-stack method-named bugs.
These cases are common: opaque reports account for 50 of the 120 benchmark bugs,
including 42 without a stack trace. The central challenge is navigating from
symptoms to causal code without direct code references.

\findingbox{Run-to-run determinism remains a central challenge}{Repeated
attempts on the same bug can produce different diagnoses, so a single
successful run can hide brittle behavior. This instability is especially
visible on weak-evidence bugs even for the strongest model.}

Diagnosis quality varies across repeated attempts on the same bug. GPT-6 Astra has
the best aggregate block coverage, but 14 of its 120 bugs are flaky under our
definition. Claude Opus 5.5 has 19 flaky bugs, GPT-5.5 has 23, GPT-5.4 has 23, GPT-5.2-Codex has 19, Kimi K2.6 has 33,
DeepSeek V4 Flash has 36, MiMo v2.5 Pro has 22, MiMo v2.5 has 27, and
HY3 has 36. The instability is most
concerning on bugs that require navigation from weak evidence. For Mini-SWE
GPT-5.5, the no-stack opaque cell has 12/42 flaky bugs (29\%) and a mean
within-bug block-coverage standard deviation of 9.7 points, while the no-stack
method-named cell has only 2/16 flaky bugs (12\%) and a 3.7-point mean standard
deviation. Mini-SWE GPT-5.2 shows the same contrast: 11/42 no-stack opaque bugs
are flaky (26\%) compared with 1/16 no-stack method-named bugs (6\%). This
variance matters for diagnosis because it suggests that reports with little
direct code evidence create brittle search trajectories rather than only
uniformly harder examples.

\findingbox{Newer models expose trade-offs rather than uniform performance
improvements}{Broader exploration can help weak-evidence cases, but agents must
remain precise when the report already provides strong evidence. Token usage
does not correlate with coverage across different models.}

The full Mini-SWE-Agent model sweep shows that stronger models do not simply
raise every category. GPT-5.5 improves aggregate block coverage from 59.5\% to
70.0\% relative to GPT-5.2-Codex, and it improves difficult weak-evidence cells:
no-stack opaque bugs improve by 17.2 percentage points.
At the same time, GPT-5.5 emits more
predicted blocks per diagnosis than GPT-5.2-Codex, 3.6 versus 2.3, and its block
precision falls from 45.5\% to 37.3\%. GPT-5.5 recovers more causal code on
opaque reports, but its broader diagnoses also include more irrelevant code.
Agents need to search beyond weak initial clues without losing precision when
the report already provides a strong local code reference.

The additional full-suite and subset experiments show the same broad lesson.
MiMo v2.5 Pro reaches 62.0\% block coverage and 73.5\% mention coverage, but
uses 2.6M tokens per run on average. MiMo v2.5 is slightly lower at 61.1\%
block coverage and 70.4\% mention coverage while using 2.3M tokens per run.
Kimi K2.6 reaches 62.0\% block coverage, but uses 2.8M tokens per run on average
and has 33 flaky bugs; DeepSeek V4 Flash roughly matches GPT-5.2-Codex block
coverage while using 3.5$\times$ the tokens and lower block precision. Gemini
3.5 Flash is more conservative: it has the highest block precision, 54.5\%, but
only 50.8\% block coverage. In \Cref{app:exploratory-subset-experiments},
DeepSeek V4 Pro gains 5.6 points while using 2.5$\times$ the GPT-5.2 tokens.
Model scaling can improve exploration, but it is not a substitute for reliable
diagnosis control.

\findingbox{Harnesses change how much code the model names, but the
core difficulty remains}{Different harnesses encourage different diagnosis
styles: Mini-SWE names more code, Codex CLI is narrower and more precise, and
OpenCode sits between them. This shifts the coverage--precision balance, but
all harnesses still struggle when reports lack direct causal code references.}

Comparing GPT-5.2-Codex across harnesses shows that the harness affects how
diagnoses are shaped. Mini-SWE-Agent reaches 59.5\% block coverage but 45.5\%
block precision, while Codex CLI reaches lower block coverage, 49.9\%, but the
highest block precision, 52.7\%. OpenCode is between them at 51.4\% block
coverage and 49.3\% block precision. The blocks/diagnosis column in \Cref{tab:eval} is
consistent with this tradeoff: Mini-SWE GPT-5.2-Codex predicts 2.3 blocks per
diagnosis on average, compared with 1.6 for Codex CLI and 1.8 for OpenCode. These
differences matter, but they are secondary to the shared failure mode: every
harness degrades sharply when the report lacks direct causal code references.

\subsection{Failure Modes and Design Implications}

The results above describe a tunnel-vision problem. Agents often
reach relevant files, but their diagnoses are broad, unstable, and especially
weak when the report does not name the causal code. We next summarize
failure modes and their implications for future harness designs. This analysis
also shows why \bench's fine-grained evidence breakdown is useful: it provides
concrete signals for diagnosing where current agents fail and what future
harnesses should improve.

\findinghdr{Failure Mode 1: Tool-call nondeterminism drives diagnosis
nondeterminism.}{Subtle differences in command choice can move runs into
different evidence scopes, so future harnesses need adaptive tools that augment
context automatically when early evidence is incomplete.}

Tool calls are a core reason why weak-evidence diagnosis becomes unstable. The
models are often good at high-level sensemaking once they are in a relevant
region, but the task requires many precise execution decisions: which keywords
to search, which directory to enter, which file to open, which symbol to follow,
and when to abandon a branch. When the report names the relevant method,
different command orders can still converge on the same local region. When the
report is opaque, small differences in search terms, directory choices, or the
first file opened can send attempts into different subsystems.

For example, in ZooKeeper-3769, node logs are available.
The core enabler for a successful diagnosis would be to identify the
\texttt{BufferUnderflowException} inside \texttt{node1.log}. We observe that
both successful and failed runs read \texttt{node1.log} early on, but successful
runs searched for \texttt{Exception} in the log, while failed runs looked at
specific line numbers or searched for election-related logs. Those reads
revealed other transient errors but missed the
\texttt{BufferUnderflowException}. Similar behavior appears in ZooKeeper-2845.
The root cause is in \texttt{ZooKeeperServer.java} under the broader
\texttt{/server} path, but the issue report strongly emphasizes quorum behavior.
Existing agents often narrow early to the \texttt{/server/quorum} subdirectory.
Once there, keyword search no longer surfaces the correct class, and the run is
steered toward a wrong but coherent diagnosis.

\noindent\textbf{Design Implication 1.} To reduce nondeterminism, harnesses should
provide tooling layers that infer the model's intent and automatically augment
the context with relevant information. For example, when an agent searches for a
keyword, static analysis can return related class definitions, symbol uses, and
upstream/downstream contexts. This makes tool execution more precise when two
commands are semantically aligned but differ subtly. Future harnesses should
not leave context construction entirely to ad hoc shell commands: they need
adaptive retrieval and context-augmentation tools that can surface adjacent
files, call paths, related symbols, and competing hypotheses when the current
search scope is under-supported.

\findinghdr{Failure Mode 2: Agents rarely revise initial hypotheses or discover
causal code beyond the initial evidence.}{When the first plausible code region
is not the causal one, agents often keep deepening that local explanation
instead of reopening the search. In our trace proxy, only 21.0\% of parseable
runs showed late recovery, and only 11.25\% of successful runs did so.}

We define \emph{late recovery} as a run whose final diagnosis files do not
appear among the first two inspected source files. Across 124 parseable
Mini-SWE-Agent and OpenCode runs with at least two inspected source files, late
recovery occurs in only 26 cases (21.0\%): 7/21 for Mini-SWE-Agent and 19/103
for OpenCode. Among successful runs, only 9/80 (11.25\%) are late recoveries.

Bugs like ZooKeeper-2845 and Ray-8889 illustrate this lock-in. Even when an
agent finds code that appears to match the reported symptom (such as the
downstream data consumer that raises the exception), it often stops at that
obvious location instead of asking what earlier lifecycle state made the symptom
possible. Trace inspections show that agents can ask relevant follow-up
questions, but whether they explore those questions is highly nondeterministic
and shaped by early exploration. This early lock-in reinforces Failure Mode 1:
because agents struggle to adapt their scope and reasoning, small tool-execution
inaccuracies can become diagnosis failures.

\noindent\textbf{Design Implication 2.} Distributed-systems debugging requires each
hypothesis path to accumulate its own evidence. Under the current single,
ever-growing context, alternatives become harder to compare. This calls for
context-management strategies beyond context compaction or sub-agent delegation,
such as switching context when pursuing different hypotheses or actively pruning
invalidated hypotheses.

\findinghdr{Failure Mode 3: Diagnosis is hard because agents lack reliable
self-correction signals before finalizing.}{Distributed-system failures often
suffer from reproducibility challenges. Agents often cannot run tests and get
quick feedback on whether the diagnosis is correct; instead, they must rely on
rigorous reasoning over source code and observed artifacts.}

Diagnosis often lacks a cheap pass/fail check for whether a proposed source
region is the true root cause. Agents often cannot tell, before finalizing, whether
the selected code region accounts for the full causal chain or whether another
hypothesis still needs to be explored. Even when a runnable environment is
available, agents can give up on testing because they prioritize reproducing
the entire incident rather than decomposing it into smaller, unit-test-style
checks. Similarly, when agents converge early, they sometimes stop for
engineering-practice reasons and lack explicit correctness specifications for
deciding whether upstream or downstream code should be blamed.

Our core insight is that bug diagnosis does not inherently lack correctness
signals. Rather, signals such as raised exceptions, crash sites, logs, and
documentation require consistent evaluation and cross-checking before they
become useful.

\noindent\textbf{Design Implication 3.} Future harnesses should actively inject
correctness signals into the agent's context, or use an evaluator sub-agent to
verify whether the proposed diagnosis fully explains the buggy behavior from
start to end, and whether the decision to blame one component is backed by
concrete evidence.

\section{Conclusion and Future Work}
\label{sec:conclusion}
We introduced \bench, a benchmark for evaluating LLM agents on code-level
root-cause diagnosis of real-world distributed system failures. Our evaluation
shows that current agents often reach relevant repository regions but struggle
to pinpoint the precise causal locations, especially when issue reports lack
direct code references or stack traces. Trace analysis highlights inconsistent
tool use and limited revision of initial hypotheses.  The evidence breakdown,
run-to-run flakiness, and failure-mode analyses together suggest that
distributed system debugging requires more than broader search or model
scaling. These findings motivate future research and agent harness designs that
support adaptive context construction, hypothesis-aware context management, and
correctness checks, so that exploration, revision, and validation
are explicit parts of agentic diagnosis workflows.

\section{Acknowledgements and Disclosures}

This work was supported by NAIRR Pilot grant 250038 with OpenAI API credits,
Jetstream2 and CloudBank resources, and in part by NSF grants CNS-2317698,
CNS-2317751, and CCF-2318937.

\bibliographystyle{plainnat}
\bibliography{bib/reference,bib/incidents}

\clearpage
\appendix
\crefalias{section}{appendix}
\crefalias{subsection}{appendix}
\section*{Appendix Overview}

The appendix provides supporting material for the benchmark construction,
evaluation protocol, and trace analysis.

\begin{itemize}[leftmargin=*]
  \item \Cref{app:benchmark-details} describes the released benchmark artifacts,
  example entries, curation process, classification labels, root-cause label
  format, annotation agreement, and threats to validity.
  \item \Cref{app:collection-pipeline-eval} summarizes the automated
  collection-pipeline evaluation against human labels.
  \item \Cref{app:prompts} gives the prompts used for diagnosis generation.
  \item \Cref{app:metric-sensitivity} reports metric-sensitivity checks,
  bootstrap confidence intervals, and leave-one-system-out results.
  \item \Cref{app:failure-mode-details} gives trace-level details behind the
  failure-mode discussion.
  \item \Cref{app:exploratory-subset-experiments} reports exploratory subset
  experiments for additional model families.
  \item \Cref{app:validity-evidence} reports the evaluator-validity audit,
  diagnosis-assisted repair pilot, and contamination analyses.
\end{itemize}

\clearpage
\section{Benchmark Details}
\label{app:benchmark-details}

\subsection{Artifact Layout}
\label{app:benchmark-artifacts}

Each benchmark instance is stored as a small directory under
\texttt{benchmark/bugs/<system>/<bug-id>/}. The release separates issue-side
evidence from gold labels so that an evaluator can instantiate the task without
looking at the fixing patch.

\begin{Verbatim}[frame=single]
benchmark/bugs/<system>/<bug-id>/
  report.json        # title, description, metadata, links, artifact pointer
  blocks.json        # gold pre-fix root-cause blocks
  artifact_dir/      # optional logs, screenshots, reproducer files, traces
\end{Verbatim}

\noindent The \texttt{report.json} file contains the original issue title and
description, project metadata, and flags such as whether the report contains
logs, stack traces, prior analysis, or attached artifacts. The \texttt{blocks.json}
file is the scoring target. Each block uses repository-local coordinates in the
pre-fix snapshot:

\begin{Verbatim}[frame=single]
{
  "file": "src/java/main/org/apache/zookeeper/server/ZooKeeperServer.java",
  "function": "ZooKeeperServer.shutdown",
  "start_line": 555,
  "end_line": 558
}
\end{Verbatim}

These coordinates are interpreted in the pre-fix checkout. Labels intentionally
exclude tests, incidental refactors, defensive hardening, and cleanup unless
those edits are necessary to explain the original failure mechanism.

\subsection{ZooKeeper-2845 Example}
\label{app:zk2845-example}

ZooKeeper-2845 is the running example used in the paper. The report describes a
data-inconsistency scenario where a transaction can be present in the on-disk
log but absent from the snapshot during leader election. The report repeatedly
mentions quorum behavior and leader election, while the gold label is in the
server lifecycle path:

\begin{Verbatim}[frame=single,breaklines=true]
Title: Data inconsistency issue due to retain database in leader election
Key symptom: txn is ahead of snapshot; a restarted server replays T1 and
diverges from the new quorum.
Gold file: src/java/main/org/apache/zookeeper/server/ZooKeeperServer.java
Gold function: ZooKeeperServer.shutdown
\end{Verbatim}

This example is useful because the issue-side narrative points strongly toward
\texttt{/server/quorum}, but the minimal causal block is in
\texttt{ZooKeeperServer.shutdown}. It therefore exercises the benchmark behavior
we care about: the agent must move from a plausible symptom-level subsystem to
the code that preserves inconsistent state across restart.

\begin{figure}[tbp]
  \centering
  \includegraphics[width=\textwidth]{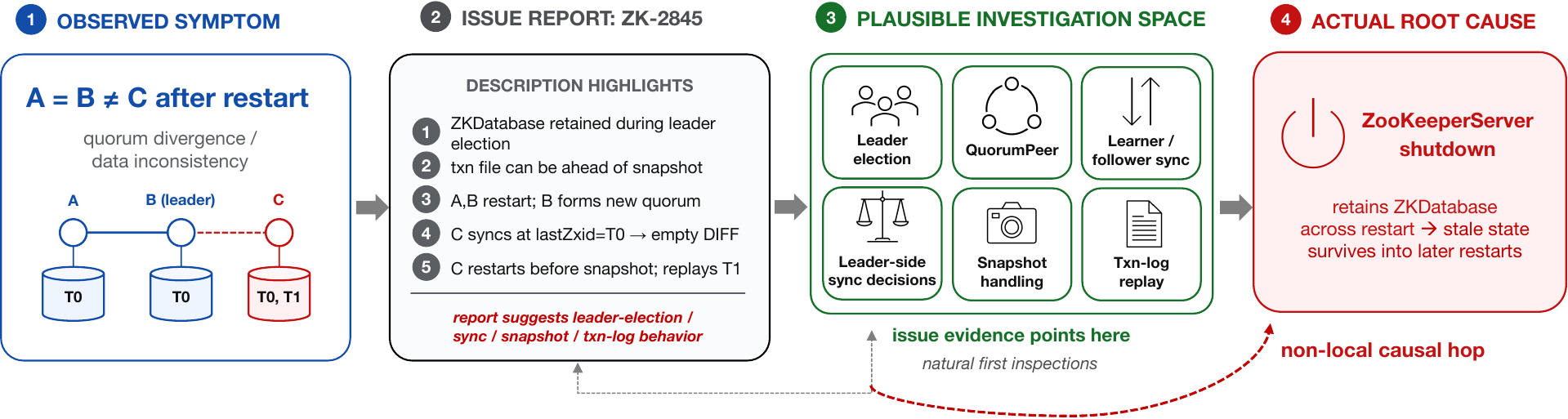}
  \caption{Causal distance in ZooKeeper-2845. The issue evidence points toward
  leader election, learner sync, snapshots, and transaction-log replay, but the
  labeled root cause is the lifecycle block in
  \texttt{ZooKeeperServer.shutdown} that retains stale \texttt{ZKDatabase}
  state across restart.}
  \label{fig:appendix-zk2845-causal-distance}
\end{figure}

\Cref{fig:appendix-zk2845-causal-distance} visualizes this distance. The
observed symptom is a quorum divergence after restart: nodes A and B agree at
state T0, while node C later replays transaction T1 from its local log. The issue
report gives useful evidence, including retained \texttt{ZKDatabase} state,
leader election, snapshot/log skew, empty DIFF sync, and txn-log replay. These
clues naturally make leader election, quorum-peer coordination, learner sync,
snapshot handling, and transaction-log replay plausible first inspection
targets. The gold root cause is different: \texttt{ZooKeeperServer.shutdown}
retains the in-memory database across restart, allowing stale state to survive
into later quorum formation. The example therefore shows why \bench measures
more than component discovery. A diagnosis must follow the non-local causal hop
from a realistic symptom-side search space to the lifecycle code that preserves
the bad state.

\begin{figure}[tbp]
  \centering
  \small
  \begin{subfigure}[t]{0.49\columnwidth}
    \centering
\begin{Verbatim}
Issue: ZOOKEEPER-2845
Title: Data inconsistency issue due to retain database in leader election

In ZOOKEEPER-2678, ZKDatabase is retained to reduce the unavailable
time during leader election. ...

But if the txn log is ahead of the snapshot, the ensemble can become
inconsistent:
1. a request is written to leader C as txn T1
2. followers A and B restart before receiving T1
3. A and B form a new quorum after restart
4. C later restarts and replays T1 from disk
5. C now contains Node N, but A and B do not
\end{Verbatim}
    \caption{Issue-side evidence}
  \end{subfigure}\hfill
  \begin{subfigure}[t]{0.49\columnwidth}
    \centering
\begin{Verbatim}
buggy_blocks:
  - file: src/java/main/org/apache/zookeeper/server/ZooKeeperServer.java
    function: ZooKeeperServer.shutdown
    lines: 528-562
  - file: src/java/main/org/apache/zookeeper/server/quorum/QuorumPeer.java
    function: QuorumPeer.getLastLoggedZxid
    lines: 926-931

metadata:
  severity: Major
  root_cause_type: Logic
  bug_complexity: Medium
  distributed: Yes
\end{Verbatim}
    \caption{Gold label}
    \label{fig:appendix-zk2845-label}
  \end{subfigure}
  \caption{Raw benchmark excerpt for \texttt{ZooKeeper-2845}.}
  \label{fig:appendix-zk2845-instance}
\end{figure}

\subsection{Curation Process}
\label{app:benchmark-curation-checks}

\bench is curated through a staged process. The goal is to retain bugs that are
realistic debugging tasks, diagnosable from issue-side evidence, and precise
enough to support source-level labels.

\paragraph{Candidate discovery.}
For each project, we collect historical public issues and apply automatic
filters for concrete fixed bugs. The filters require closed or resolved issue
status, an associated fixing change, a bug-like issue type or resolution, and an
informative report. When project metadata is available, we prioritize
high-impact user-facing issues, such as Major, Critical, or Blocker bugs on
stable releases. We exclude feature requests, documentation-only issues,
test-only issues, and reports whose description is too sparse to support
diagnosis.

\paragraph{Fix-level filtering.}
The fixing change is used only for curation and labeling, not as agent input.
We remove broad refactorings, formatting-only changes, dependency updates, and
patches dominated by unrelated cleanup. As a practical filter, we exclude fixes
with more than 100 changed source lines unless the causal span is clearly
separable from auxiliary edits.

\begin{table}[tbp]
  \centering
  \small
  \caption{Example attrition in the ZooKeeper curation slice.}
  \label{tab:appendix-zookeeper-curation}
  \begin{tabular}{lrp{0.48\columnwidth}}
    \toprule
    Stage & Issues & Selection step \\
    \midrule
    Public issue pool & 4,789 & Historical ZooKeeper issues considered by the
    collection pipeline. \\
    Automatic candidates & 99 & Closed fixed bugs that pass metadata, report,
    and fixing-change filters. \\
    Manual inspection sample & 50 & Candidate issues sampled for human audit and
    root-cause labelability checks. \\
    Released ZooKeeper incidents & 18 & Accepted incidents with diagnosable
    issue-side context and reconciled source-level labels. \\
    \bottomrule
  \end{tabular}
\end{table}

\paragraph{Manual audit.}
Annotators check that an issue is diagnosable from issue-side context alone.
Issue reports and attached artifacts are used as the diagnosis context, while
fixing commits are used only for human annotation. Later issue-thread comments
that reveal the root cause or patch strategy are removed or ignored. This keeps
the task focused on diagnosis from incident evidence rather than retrieval of
the final answer from post-fix discussion.

\paragraph{Root-cause reconciliation.}
Each accepted issue is independently inspected by three annotators. The
annotators derive candidate pre-fix root-cause blocks from the fixing change and
surrounding code context, exclude tests and incidental cleanup, and reconcile
disagreements into a final block set. The reconciled labels are the only labels
used for benchmark scoring.

The benchmark also records characterization metadata for each accepted issue.
These metadata fields are not scoring targets and are not provided as hints to
agents. They support the analysis in the paper by letting us break down results
by textual code anchors, crash-to-fix distance, issue-side evidence, and
root-cause structure.

\subsection{Classification Axes}
\label{app:benchmark-classifications}

The benchmark characterization uses two gold-aware axes. These classifications
are used only for analysis and figure breakdowns; they are not provided to
agents at evaluation time.

\begin{table}[tbp]
  \centering
  \small
  \caption{Textual code-reference categories used in the evaluation breakdown.}
  \label{tab:appendix-textual-categories}
  \begin{tabular}{lll}
    \toprule
    Category & Definition & Count \\
    \midrule
    Method-named & A non-generic gold function name appears in the issue text & 38 \\
    File/class-named & A gold file stem or class name appears, but no gold method & 32 \\
    Opaque & No extracted gold method, file, or class identifier appears & 50 \\
    \bottomrule
  \end{tabular}
\end{table}

\begin{table}[tbp]
  \centering
  \small
  \caption{Crash-to-fix categories used in the evaluation breakdown.}
  \label{tab:appendix-crash-categories}
  \begin{tabular}{lll}
    \toprule
    Category & Definition & Count \\
    \midrule
    Co-located & First project-internal stack frame is in a gold file & 13 \\
    Adjacent & Crash frame and gold file are in nearby package structure & 6 \\
    Cross-component & Stack trace exists, but the first internal frame is elsewhere & 26 \\
    No stack trace & No usable project-internal stack frame is found & 75 \\
    \bottomrule
  \end{tabular}
\end{table}

For the main category figure, we merge co-located and adjacent into a
\emph{local/adjacent} row because the adjacent cell is small. The resulting
3-by-3 table contains 42 no-stack opaque bugs, 16 no-stack method-named bugs,
17 no-stack file/class-named bugs, 7 cross-component opaque bugs, 12
cross-component method-named bugs, 7 cross-component file/class-named bugs, 1
local/adjacent opaque bug, 10 local/adjacent method-named bugs, and 8
local/adjacent file/class-named bugs.

\begin{table}[tbp]
  \centering
  \small
  \caption{Issue-side signal fields extracted for benchmark characterization.}
  \label{tab:appendix-issue-signals}
  \begin{tabular}{lr}
    \toprule
    Signal present in report or attached evidence & Incidents \\
    \midrule
    At least one code identifier & 96 \\
    Prior analysis or causal explanation & 37 \\
    Stack trace & 28 \\
    Logs & 20 \\
    Cross-linked issue reference & 22 \\
    Attached artifacts & 7 \\
    \bottomrule
  \end{tabular}
\end{table}

We also assign a coarse root-cause type to each incident for benchmark
characterization. These types are descriptive and are not used by the evaluator
or exposed to agents. They give a high-level view rather than an exhaustive
failure taxonomy; finer distinctions such as state management, protocol
interaction, recovery, configuration, and multi-component coordination would
require separate annotation.

\begin{table}[tbp]
  \centering
  \small
  \caption{Coarse root-cause categories in the full 120-incident benchmark.}
  \label{tab:appendix-root-cause-types}
  \begin{tabular}{lr}
    \toprule
    Root-cause type & Incidents \\
    \midrule
    Logic & 56 \\
    Concurrency & 28 \\
    Error handling & 26 \\
    Performance & 4 \\
    Resource leak & 3 \\
    Communication & 2 \\
    Dependency & 1 \\
    \bottomrule
  \end{tabular}
\end{table}

\subsection{Label Structure}
\label{app:benchmark-label-structure}

\bench contains 219 gold buggy blocks across 120 incidents. Multi-block labels
are common: 55 incidents (45.8\%) require more than one block, and 29 incidents
(24.2\%) span multiple files. The median labeled block length is four lines. This
is why the paper reports file/function/line metrics but uses block coverage as
the primary diagnosis metric: useful diagnoses must reach the right region
without being overly sensitive to exact line-boundary choices.

\paragraph{Annotation agreement.}
\label{app:annotation-agreement}
All 120 incidents followed the same three-annotator labeling and reconciliation
workflow. For 51 incidents, we retained the separate annotation sheets from
before adjudication, allowing us to measure agreement among the independent
labels. For the other 69 incidents, disagreements were resolved through
discussion without retaining separate pre-adjudication sheets. The agreement
statistics below therefore describe the 51 retained cases.

We assess agreement at file, normalized-function, and block granularity.
\Cref{tab:annotation-agreement} reports Krippendorff's $\alpha$ and pairwise
Jaccard similarity at the first two levels, and block coverage and F1 under
the benchmark's five-line-tolerant block criterion. Annotators agree strongly
on the causal files and functions. Differences in exact line-span boundaries
and in how annotators split a causal region into blocks reduce exact-span
agreement, while agreement under the block criterion remains high.

\begin{table}[htbp]
  \centering\small
  \caption{Pre-adjudication annotation agreement on the 51 incidents with
  retained individual annotation sheets.}
  \label{tab:annotation-agreement}
  \begin{tabular}{@{}llr@{}}\toprule
  Granularity & Measure & Agreement \\
  \midrule
  File & Krippendorff's $\alpha$ & 0.919 \\
       & Pairwise Jaccard & 0.902 \\
  Normalized function & Krippendorff's $\alpha$ & 0.866 \\
       & Pairwise Jaccard & 0.837 \\
  Block (five-line tolerance) & Coverage & 0.861 \\
       & F1 & 0.848 \\
  \bottomrule\end{tabular}
\end{table}

\subsection{Threats to Validity}
\label{app:benchmark-validity}

\paragraph{Internal validity.}
The main internal risks are that the benchmark labels or evaluator may not
faithfully measure the diagnosis behavior we intend to study.

\textbf{Label validity.} Distributed-system fixes can include upstream state
transitions, downstream guards, invariant checks, tests, and auxiliary cleanup.
We mitigate this by deriving labels from the historical fixing change and
surrounding code context, having three annotators independently inspect each
issue, reconciling disagreements into a final block set, and labeling only the
minimal pre-fix code that embodied the failure mechanism. Tests, helper
refactors, defensive hardening, and incidental cleanup are excluded unless they
are necessary to the causal explanation. On the 51 incidents with retained
independent annotation sheets, agreement is high at the file and function levels
(\Cref{app:annotation-agreement}).

\textbf{Evaluation sensitivity.} A distributed-system failure may be mitigated
or repaired at several locations, but the causal chain that produces it is
specific. In ZooKeeper-2845, for example, one could add defensive handling later
in the synchronization or recovery path. Such a change might reduce the
observed inconsistency, but it would not identify the underlying bug: the
retained database and shutdown/restart behavior let the state used during
synchronization diverge from the transaction log. \bench therefore labels the
causal code identified through issue resolution and expert annotation, rather
than every location where a symptom could be detected, masked, or repaired; it
does not claim that the historical patch is the only valid fix. A diagnosis that
blames a different plausible intervention point can still receive low block
coverage. The paper therefore reports file, function, line, and block metrics,
together with mention coverage, which records whether the causal explanation
refers to the labeled functions even when the selected blocks differ.
\Cref{app:evaluator-validation} shows that such alternatives explain few
zero-coverage diagnoses and that higher-coverage diagnoses guide repair more
effectively. Future evaluators could combine multi-reference annotations with
semantic and program-analysis-based validation to better capture equivalent
causal explanations.

\textbf{Benchmark leakage.} \bench is built from public user-reported failures.
This makes the benchmark realistic and reproducible, but it also means that
model pretraining exposure cannot be ruled out completely. Public benchmark
identifiers and future fix information can contaminate an evaluation if agents
retrieve them rather than diagnose from allowed evidence. We mitigate direct
answer leakage during benchmark construction by removing issue-thread discussion
that reveals the root cause or patch strategy, anonymizing workspace names, and
removing commits after the task snapshot. The leakage guard~\autoref{fig:teaser}
scans visible reasoning and tool inputs/outputs for the original issue identifiers
withheld from the agent. This check targets direct retrieval of historical
answers, including searches for future fixing commits. Flagged attempts are
discarded and rerun, and the team investigates residual identifiers or accessible
fix information in the prepared environment. Because detection depends on
explicit identifier matches, implicit recall requires separate evaluation.
\Cref{app:contamination-evidence} reports guard-hit counts, closed-book probes,
and correlations with public-exposure proxies.

\paragraph{External validity.}
The main external risk is whether conclusions from \bench generalize beyond this
benchmark sample.

\textbf{Selection bias.} \bench intentionally focuses on closed, high-impact
failures in production distributed systems whose fixes support precise
source-level labeling. This selection makes the task realistic and scoreable,
but it excludes failures whose fixes are dominated by broad refactorings,
configuration changes, test-only changes, or ambiguous cleanup. The benchmark
should therefore be interpreted as covering code-level diagnosis for labelable
distributed-system bugs rather than all incident-response work.

We mitigate this risk with a principled, staged selection process. For each
project, we first collect historical public issues and apply automatic filters
for concrete fixed bugs: closed or resolved status, an associated fixing change,
high-priority or user-facing failures when priority is available, sufficiently
informative reports, and non-test/non-documentation components. We then filter
by the fixing change itself, excluding broad refactorings and patches dominated
by unrelated cleanup, and using a practical 100 changed-source-line cutoff
unless the causal span is clearly separable from auxiliary edits. The ZooKeeper
attrition example in \Cref{tab:appendix-zookeeper-curation} illustrates the
scale of this filtering process. Finally, annotators manually audit the
survivors to confirm that the issue-side context is diagnosable without
post-report discussion, that the report does not reveal the root cause or patch
strategy, and that the fixing change maps to precise pre-fix root-cause blocks.
Each accepted issue is then independently annotated by three annotators and
reconciled into the final label set. This process does not eliminate selection
bias, but it makes the benchmark population the result of consistent filters and
explicit audits rather than convenience sampling.

\section{LLM-Assisted Collection Pipeline}
\label{app:collection-pipeline-eval}
Benchmark curation requires human review for two reasons: identifying true bug
reports while excluding feature requests, and identifying the source blocks that
encode each root cause. We developed an LLM-assisted pipeline for the second
bottleneck. The pipeline starts from a bug-fixing commit, fetches the diff
automatically, and prompts an LLM to output only the relevant root-cause blocks.
Preliminary checks show a non-trivial gap between LLM-produced labels and human
labels, but we still see potential in LLM-assisted automation pipelines. Below,
we report a small-subset comparison between this LLM-assisted pipeline, a
non-LLM extraction pipeline, and the final human labels. The first bottleneck is
also important, but determining whether a bug is diagnosable from the available
information is essentially asking LLM agents to diagnose the bug itself, which
is the task we are trying to test in the first place. We will explore fully
automating \bench's curation process while maintaining benchmark quality, with
more fine-grained bug classification based on information availability.

We evaluate two benchmark-construction variants against the human labels used in
\bench. The LLM-assisted variant starts from the bug-fixing diff and prompts an
LLM to select the source blocks that correspond to the root cause, excluding
tests, helper refactors, and other auxiliary edits. The non-LLM variant is a
deterministic extraction pipeline that derives candidate labels from the fixing
change using source-level parsing and filtering, without asking an LLM to judge
semantic relevance. Both variants address the second curation bottleneck from
\Cref{sec:benchmark}: separating root-cause code blocks from other edits in a
known fixing change.

The LLM-assisted variant is available only on a small subset, so the table below
reports both variants on the same 12 bugs completed by the LLM-assisted
pipeline. The non-LLM pipeline covers additional bugs, but those are omitted
from this comparison to keep the subset fixed.

\begin{table}[tbp]
  \centering
  \small
  \caption{Collection-pipeline hit-rate comparison on the 12-bug shared subset.
  Scores compare generated labels with human labels.}
  \label{tab:collection-pipeline-shared-summary}
  \begin{tabular}{lrrrrr}
    \toprule
    Construction method & GT blocks & Pred. blocks & Recall & Precision & F1 \\
    \midrule
    LLM-assisted label generation & 22 & 15 & 54.5\% & 73.3\% & 62.6\% \\
    Non-LLM extraction pipeline & 22 & 52 & 77.3\% & 34.6\% & 47.8\% \\
    \bottomrule
  \end{tabular}
\end{table}

\Cref{tab:collection-pipeline-shared-summary} shows a coverage--precision
trade-off. On the shared subset, the non-LLM pipeline recovers more
human-labeled blocks, but it does so by emitting many more candidate labels. The
LLM-assisted pipeline is more conservative: it predicts 15 blocks compared with
52 for the non-LLM pipeline, yielding higher precision but lower recall. This
suggests that LLM-assisted label extraction is a promising direction for
scaling benchmark construction, but the remaining recall gap shows that it is
not yet a substitute for human label review and motivates future work on more
reliable automated curation.

\section{Evaluation Prompts and Output Contract}
\label{app:prompts}

All evaluated harnesses receive the same diagnosis task: inspect the pre-fix
repository and issue-side evidence, identify the pre-fix code blocks that encode
the root cause, and emit a structured diagnosis. The exact interaction syntax
differs by harness, but the task-level constraints are shared:

\begin{itemize}
  \item the agent should diagnose rather than patch the bug;
  \item the agent should not modify repository source files;
  \item lightweight static analysis is preferred over long-running integration
  tests or networked services;
  \item the final answer must be valid JSON in the required schema; and
  \item each completed bug/configuration pair contributes three valid diagnoses.
\end{itemize}

\subsection{Mini-SWE-Agent Prompt Excerpt}

The Mini-SWE-Agent experiments use a diagnosis-only system prompt and an
instance prompt that instructs the model to write \texttt{diagnosis.json}. The
following excerpt shows the task-level content, omitting harness-specific
formatting examples for shell actions.

\begin{Verbatim}[frame=single,breaklines=true]
You are a helpful assistant that can interact with a computer through bash
tool calls.

Your job is diagnosis only. Do not modify the codebase and do not attempt to
produce a fix.

Please debug the root cause of this distributed-systems issue:

{{ issue_title_and_description }}

Produce a code-level diagnosis of the bug. Your final output is a JSON file
named diagnosis.json. Do not include Markdown in the file.

Requirements:
- Do not modify repository source files.
- Use the repository, issue text, stack traces, and attached artifacts to reason
  about the root cause.
- You may inspect code and run lightweight local commands.
- Avoid long-running integration tests or networked services.
- Keep diagnosis.json as pure JSON with exactly the schema below.
- Before finishing, validate diagnosis.json with the provided validator.
\end{Verbatim}

\subsection{Codex CLI and OpenCode Prompt Excerpt}

For Codex CLI and OpenCode, the prompt is expressed in the harness' native task
format, but preserves the same diagnosis-only contract. The following excerpt is
representative of the task message:

\begin{Verbatim}[frame=single,breaklines=true]
You are auditing a historical bug report for a distributed system.
Your task is to inspect the repository, understand the failure scenario, and
identify the exact code blocks responsible.

Instructions:
- Work only inside the current Git repository, already checked out at the target
  pre-fix commit.
- Use lightweight static analysis such as search, grep, and reading code.
- Avoid running integration tests unless necessary.
- When you reach a conclusion, provide a JSON object that follows the supplied
  schema. Do not embed Markdown or extra commentary.
- Each buggy block must have an accurate file, function, start_line, and
  end_line.
- Include a concise bug_cause_description tying everything together.
\end{Verbatim}

\subsection{Diagnosis Schema}

The required output schema is intentionally small. The evaluator ignores
optional narrative fields for scoring and compares the predicted block
coordinates against the gold labels at file, function, line, and block
granularity.

\begin{Verbatim}[frame=single,breaklines=true]
{
  "buggy_blocks": [
    {
      "file": "relative/path/to/file",
      "function": "fully.qualified.FunctionName",
      "start_line": 123,
      "end_line": 130,
      "bug_contribution": "How this block contributes to the bug.",
      "patch": "Optional minimal patch sketch or explanation."
    }
  ],
  "bug_cause_description": "Overall root-cause explanation."
}
\end{Verbatim}

The \texttt{patch} field is present for compatibility with some harness outputs,
but patch quality is not scored in this release. The scoring pipeline uses only
the predicted source coordinates and, for mention coverage, the natural-language
function references in \texttt{bug\_cause\_description}.

\section{Coverage Metric Sensitivity}
\label{app:metric-sensitivity}

The main paper uses block coverage as the primary localization metric because
it asks whether the agent reached each labeled causal region while being less
sensitive to exact line-boundary choices. Line coverage is stricter: a diagnosis
can reach the correct block but receive lower line coverage if the predicted span
is shifted, shorter, or longer than the human label. We therefore treat line
coverage as a sensitivity check rather than the main story.

\begin{figure}[tbp]
  \centering
  \includegraphics[width=\textwidth]{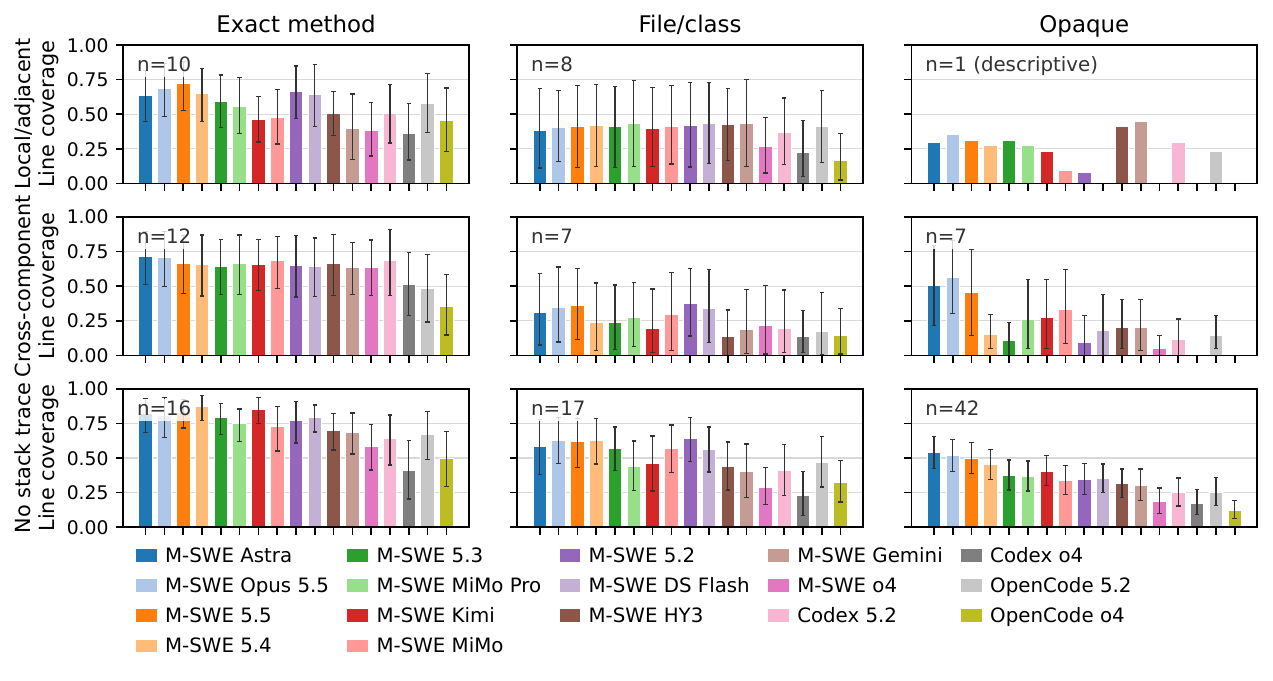}
  \caption{Line coverage by incident evidence. Columns split bugs by
  whether the issue text contains an exact gold method, only a gold file/class
  anchor, or no direct textual anchor. Rows split bugs by whether the first
  crash stack frame is local or adjacent to the fix, in another component, or
  absent. Line coverage is stricter than block coverage and measures overlap
  with labeled causal lines. Panel counts give bugs per cell; whiskers show
  95\% bug-bootstrap CIs (none for the single-bug cell).}
  \label{fig:eval-evidence-line-coverage}
\end{figure}

\Cref{fig:eval-evidence-line-coverage} shows that the qualitative pattern
from the main block-coverage figure remains visible under the stricter metric:
opaque and no-stack bugs remain much harder than method-named bugs. The absolute
values are lower because line overlap penalizes span-boundary mismatch.

\begin{figure}[t]
  \centering
  \includegraphics[width=0.78\textwidth]{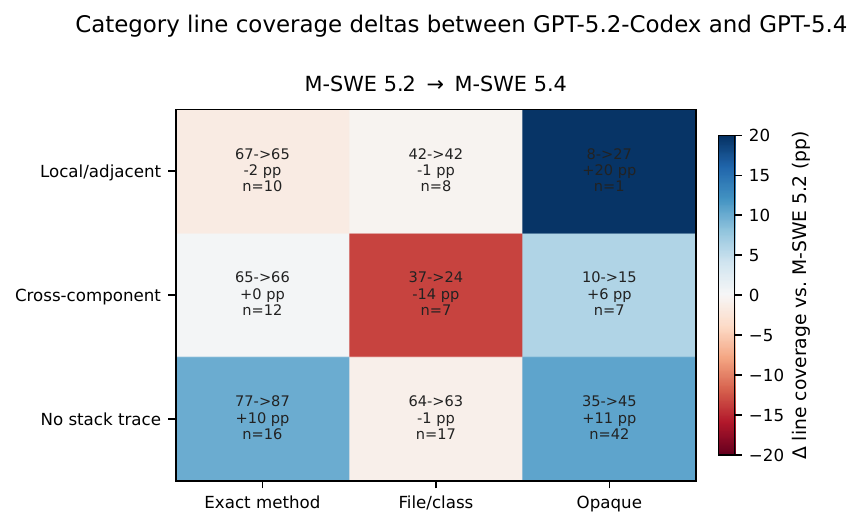}
  \caption{Paired line-coverage deltas between Mini-SWE GPT-5.2-Codex and
  Mini-SWE GPT-5.4 by evidence category. Each cell reports absolute
  GPT-5.2$\rightarrow$GPT-5.4 line coverage, the paired delta in percentage
  points, and the number of bugs.}
  \label{fig:gpt54-vs-gpt52-line-category-deltas}
\end{figure}

\Cref{fig:gpt54-vs-gpt52-line-category-deltas} provides the paired GPT-5.4
versus GPT-5.2 sensitivity view. The line metric shows the same caution as the
block metric: newer models improve several weak-evidence cells, but the changes
are not uniform across categories.

\subsection{Aggregation and Statistical Uncertainty}
\label{app:eval-uncertainty}

For metric scores $s_{ir}$ on bug $i$ and attempt $r$, we first compute
$\bar{s}_i=R_i^{-1}\sum_{r=1}^{R_i}s_{ir}$ and then report
$\hat{\mu}=n^{-1}\sum_{i=1}^{n}\bar{s}_i$. Each included bug has $R_i=3$ selected attempts, and bugs have equal weight. Each bootstrap replicate
samples $n$ bug IDs with replacement and averages their $\bar{s}_i$ values.
The 2.5th and 97.5th percentiles of 10,000 replicate means form the reported
95\% interval. We use sorted bug IDs, NumPy's PCG64 generator with seed 2026,
and linear interpolation for quantiles. Category intervals use the same
procedure within each category. A cell with fewer than two bugs has no CI.

These intervals treat bugs as independent draws from the incidents the
benchmark represents. They capture sensitivity to which bugs are sampled, not
rerun variability on a fixed bug, which we report separately as within-bug
standard deviations and flaky-bug counts. Because bugs from the same system may
be correlated, \Cref{app:eval-system-sensitivity} also reports per-system and
leave-one-system-out results.

\paragraph{Paired differences.}
For configurations $A$ and $B$, we compute one difference per bug,
$d_i=\bar{s}_{i,A}-\bar{s}_{i,B}$, and estimate
$\hat{\Delta}=n^{-1}\sum_i d_i$. Each replicate samples the same $n$ bug IDs
for both configurations, equivalently resampling the $d_i$ values. We use the
same 10,000-replicate percentile procedure, seed, and quantile convention as
above. Pairing is by bug ID, not by attempt index; the three selected attempts are
averaged within each bug before resampling.

\Cref{tab:eval-paired-confidence} reports block-coverage differences for the twelve
other Mini-SWE configurations against GPT-5.2-Codex, GPT-5.5 against GPT-5.4,
and the four matched-model Mini-SWE versus Codex CLI/OpenCode comparisons.
These intervals are unadjusted for multiple comparisons. Pairing makes model
comparisons more sensitive than comparing separate intervals: the individual
CIs of Mini-SWE GPT-5.5 and GPT-5.2-Codex overlap, yet GPT-5.5's 10.50-point
gain has a paired CI of $[6.34, 15.12]$.

\begin{table}[tbp]
\centering\scriptsize
\caption{Mean scores and 95\% bug-level bootstrap confidence intervals across 120 common bugs, in percent.}
\label{tab:eval-confidence-intervals}
\resizebox{\textwidth}{!}{%
\begin{tabular}{@{}lrrrrrr@{}}\toprule
Configuration & File & Function & Line & Block & Block precision & Mention \\\midrule
MiniSWE GPT-6 Astra & 86.1 [80.8, 91.0] & 59.8 [51.9, 67.8] & 58.2 [51.4, 65.2] & 72.6 [65.7, 79.3] & 46.0 [40.1, 52.1] & 83.3 [77.4, 88.8] \\
MiniSWE Claude Opus 5.5 & 84.5 [79.1, 89.6] & 61.8 [54.3, 69.2] & 58.9 [52.2, 65.6] & 71.8 [65.2, 78.2] & 49.0 [43.2, 54.7] & 82.1 [76.2, 87.5] \\
MiniSWE GPT-5.5 & 86.2 [80.8, 91.0] & 57.4 [49.5, 65.1] & 57.7 [50.9, 64.5] & 70.0 [63.1, 76.7] & 37.3 [32.4, 42.2] & 76.0 [69.4, 82.2] \\
MiniSWE GPT-5.4 & 81.0 [75.3, 86.2] & 54.6 [46.9, 62.3] & 53.6 [46.8, 60.4] & 66.6 [59.7, 73.4] & 41.6 [36.5, 46.8] & 70.4 [63.6, 76.8] \\
MiniSWE GPT-5.3-Codex & 79.9 [73.6, 85.7] & 54.1 [46.1, 61.9] & 48.1 [41.4, 54.9] & 63.3 [55.9, 70.4] & 42.7 [37.2, 48.2] & 69.9 [62.9, 76.6] \\
MiniSWE MiMo v2.5 Pro & 80.5 [74.3, 86.2] & 60.6 [53.1, 68.1] & 46.7 [39.7, 53.5] & 62.0 [54.5, 69.3] & 40.2 [34.5, 46.0] & 73.5 [66.6, 80.0] \\
MiniSWE Kimi K2.6 & 79.2 [73.3, 84.7] & 51.9 [44.6, 59.2] & 48.0 [41.2, 55.0] & 62.0 [54.8, 69.0] & 49.2 [42.7, 55.8] & 71.4 [64.6, 77.8] \\
MiniSWE MiMo v2.5 & 79.7 [73.5, 85.4] & 54.9 [47.4, 62.4] & 46.9 [40.2, 53.9] & 61.1 [53.8, 68.2] & 38.7 [33.3, 44.3] & 70.4 [63.5, 77.1] \\
MiniSWE GPT-5.2-Codex & 76.7 [69.9, 83.0] & 49.6 [41.5, 57.6] & 49.2 [42.0, 56.3] & 59.5 [51.7, 67.1] & 45.5 [39.0, 51.8] & 64.4 [56.9, 71.7] \\
MiniSWE DeepSeek V4 Flash & 77.4 [71.3, 83.0] & 50.1 [42.3, 57.9] & 48.6 [41.7, 55.5] & 59.2 [52.0, 66.2] & 34.6 [29.5, 39.7] & 68.7 [61.7, 75.4] \\
MiniSWE HY3 & 74.1 [67.3, 80.8] & 49.8 [42.3, 57.5] & 42.5 [35.8, 49.1] & 54.8 [47.4, 62.2] & 41.3 [34.7, 47.9] & 65.3 [58.1, 72.3] \\
MiniSWE Gemini 3.5 Flash & 67.7 [60.2, 74.6] & 44.4 [36.7, 52.2] & 40.7 [33.6, 47.8] & 50.8 [43.1, 58.3] & 54.5 [46.4, 62.2] & 59.7 [52.4, 67.0] \\
MiniSWE o4-mini & 64.2 [56.7, 71.3] & 36.3 [29.4, 43.7] & 31.3 [25.1, 37.6] & 42.1 [34.9, 49.6] & 37.3 [30.6, 43.9] & 50.7 [43.2, 58.1] \\
Codex CLI GPT-5.2-Codex & 65.7 [58.2, 73.0] & 47.7 [39.8, 55.7] & 38.7 [31.7, 45.9] & 49.9 [42.0, 57.7] & 52.7 [44.6, 60.6] & 60.2 [52.7, 67.5] \\
Codex CLI o4-mini & 56.5 [48.9, 63.9] & 37.5 [30.2, 44.9] & 25.3 [19.1, 31.7] & 35.9 [28.6, 43.5] & 35.5 [28.0, 43.1] & 52.0 [44.4, 59.5] \\
OpenCode GPT-5.2-Codex & 69.4 [62.3, 76.2] & 45.3 [37.5, 53.1] & 39.0 [32.1, 46.2] & 51.4 [43.7, 59.1] & 49.3 [41.8, 56.8] & 58.6 [51.1, 66.0] \\
OpenCode o4-mini & 55.0 [47.3, 62.6] & 31.3 [24.2, 38.5] & 24.8 [19.0, 30.8] & 35.1 [28.0, 42.2] & 40.5 [32.8, 48.2] & 45.4 [37.7, 53.0] \\
\bottomrule\end{tabular}}
\end{table}

\begin{table}[tbp]
\centering\small
\caption{Paired block-coverage differences on the 120 common bugs, in percentage points (candidate minus reference). Intervals are unadjusted pairwise 95\% bug-bootstrap CIs; they are not simultaneous leaderboard bounds.}
\label{tab:eval-paired-confidence}
\setlength{\tabcolsep}{4pt}
\begin{tabular}{@{}llrr@{}}\toprule
Candidate & Reference & Difference & 95\% CI \\\midrule
MiniSWE GPT-6 Astra & MiniSWE GPT-5.2-Codex & $+13.13$ & $[+7.43, +19.35]$ \\
MiniSWE Claude Opus 5.5 & MiniSWE GPT-5.2-Codex & $+12.32$ & $[+7.50, +17.58]$ \\
MiniSWE GPT-5.5 & MiniSWE GPT-5.2-Codex & $+10.50$ & $[+6.34, +15.12]$ \\
MiniSWE GPT-5.4 & MiniSWE GPT-5.2-Codex & $+7.04$ & $[+3.33, +11.03]$ \\
MiniSWE GPT-5.3-Codex & MiniSWE GPT-5.2-Codex & $+3.74$ & $[+0.50, +7.36]$ \\
MiniSWE MiMo v2.5 Pro & MiniSWE GPT-5.2-Codex & $+2.50$ & $[-2.47, +7.58]$ \\
MiniSWE Kimi K2.6 & MiniSWE GPT-5.2-Codex & $+2.49$ & $[-2.44, +7.62]$ \\
MiniSWE MiMo v2.5 & MiniSWE GPT-5.2-Codex & $+1.58$ & $[-2.74, +6.23]$ \\
MiniSWE DeepSeek V4 Flash & MiniSWE GPT-5.2-Codex & $-0.36$ & $[-4.70, +3.82]$ \\
MiniSWE HY3 & MiniSWE GPT-5.2-Codex & $-4.69$ & $[-9.65, +0.33]$ \\
MiniSWE Gemini 3.5 Flash & MiniSWE GPT-5.2-Codex & $-8.71$ & $[-14.83, -2.50]$ \\
MiniSWE o4-mini & MiniSWE GPT-5.2-Codex & $-17.38$ & $[-22.63, -12.37]$ \\
MiniSWE GPT-5.5 & MiniSWE GPT-5.4 & $+3.46$ & $[-0.04, +7.24]$ \\
MiniSWE GPT-5.2-Codex & Codex CLI GPT-5.2-Codex & $+9.63$ & $[+5.21, +14.41]$ \\
MiniSWE GPT-5.2-Codex & OpenCode GPT-5.2-Codex & $+8.09$ & $[+3.70, +12.68]$ \\
MiniSWE o4-mini & Codex CLI o4-mini & $+6.21$ & $[+1.67, +10.77]$ \\
MiniSWE o4-mini & OpenCode o4-mini & $+7.06$ & $[+1.86, +12.10]$ \\
\bottomrule\end{tabular}
\end{table}

\clearpage
\subsection{System Composition and Leave-One-System-Out Sensitivity}
\label{app:eval-system-sensitivity}

We examine whether the aggregate result is dominated by one system using
Mini-SWE-Agent with GPT-5.4. \Cref{tab:eval-per-system-confidence} reports block coverage and
95\% CIs for all 11 systems. Each bug contributes the mean of its three
selected attempts; system-specific intervals resample bugs within that system
using the procedure in \Cref{app:eval-uncertainty}.

For each leave-one-system-out comparison, we remove every bug from one system
and recompute the file- and block-coverage means over the remaining bugs.
We retain equal weight per bug and bootstrap the remaining bugs for each
interval; we do not average the system-level means.
\Cref{tab:eval-leave-one-system-out} gives all 11 exclusions and the full-cohort
reference. Removing a system changes block coverage by at most 2.1 percentage
points, from a full-cohort mean of 66.6\%. The file-to-block coverage gap remains
between 13.1 and 15.3 points under every exclusion. The difficulty of precise
causal-code localization therefore persists without any single system.

\begin{table}[htbp]\centering\small
\caption{MiniSWE GPT-5.4 block coverage by system. Scores and 95\% bug-bootstrap CIs are percentages; each bug has three selected attempts.}
\label{tab:eval-per-system-confidence}
\begin{tabular}{@{}lrr@{}}\toprule
System & Bugs & Block coverage [95\% CI] \\\midrule
Overall & 120 & 66.6 [59.7, 73.4] \\
Elasticsearch & 19 & 70.1 [53.3, 85.4] \\
etcd & 8 & 78.1 [61.5, 92.7] \\
HBase & 5 & 40.0 [13.3, 73.3] \\
HDFS & 10 & 52.7 [26.0, 79.3] \\
Ignite & 3 & 61.1 [0.0, 100.0] \\
Ignite 3 & 7 & 64.3 [35.7, 88.1] \\
Kafka & 8 & 79.2 [58.3, 93.8] \\
Kubernetes & 10 & 81.7 [60.0, 100.0] \\
Ray & 21 & 56.7 [41.1, 72.2] \\
TiKV & 11 & 55.6 [29.3, 80.8] \\
ZooKeeper & 18 & 78.7 [58.3, 95.4] \\
\bottomrule\end{tabular}\end{table}
\begin{table}[htbp]\centering\small
\caption{Leave-one-system-out results for MiniSWE GPT-5.4. Coverage and CIs are percentages. $\Delta$ is block coverage minus the full-cohort mean; the file--block gap and $\Delta$ are in percentage points. Each remaining bug retains equal weight.}
\label{tab:eval-leave-one-system-out}
\setlength{\tabcolsep}{4pt}
\begin{tabular}{@{}lrrrrr@{}}\toprule
Omitted system & Bugs left & File & Block [95\% CI] & $\Delta$ & File--block gap \\\midrule
None & 120 & 81.0 & 66.6 [59.7, 73.4] & $+0.0$ & 14.5 \\
Elasticsearch & 101 & 79.9 & 65.9 [58.1, 73.3] & $-0.7$ & 14.0 \\
etcd & 112 & 80.7 & 65.7 [58.3, 72.8] & $-0.8$ & 15.0 \\
HBase & 115 & 82.1 & 67.7 [60.4, 74.5] & $+1.2$ & 14.4 \\
HDFS & 110 & 80.9 & 67.8 [60.8, 74.7] & $+1.3$ & 13.1 \\
Ignite & 117 & 81.0 & 66.7 [59.7, 73.5] & $+0.1$ & 14.3 \\
Ignite 3 & 113 & 81.5 & 66.7 [59.5, 73.6] & $+0.1$ & 14.8 \\
Kafka & 112 & 80.6 & 65.6 [58.4, 72.7] & $-0.9$ & 14.9 \\
Kubernetes & 110 & 80.2 & 65.2 [58.1, 72.3] & $-1.4$ & 15.1 \\
Ray & 99 & 83.6 & 68.6 [60.8, 76.1] & $+2.1$ & 15.0 \\
TiKV & 109 & 81.4 & 67.7 [60.4, 74.6] & $+1.1$ & 13.8 \\
ZooKeeper & 102 & 79.7 & 64.4 [56.8, 71.5] & $-2.1$ & 15.3 \\
\bottomrule\end{tabular}\end{table}

\clearpage
\subsection{Attempt Selection and Common Evaluation Cohort}
\label{app:eval-attempt-counts}

For each bug/configuration pair, we sort attempts by numeric run ID and retain
the first three with a schema-valid \texttt{diagnosis.json} and completed,
matching scores. Attempts missing these artifacts do not occupy a slot; later
completed attempts can replace them. Selection does not depend on the score:
completed incorrect diagnoses, including schema-valid empty predictions, remain
in the evaluation. Additional valid attempts after the first three are unused.

The primary comparison uses all 120 benchmark incidents across 17
configurations: 13 Mini-SWE model configurations, plus two models on each of
Codex CLI and OpenCode. The three-attempt protocol gives
$120\times17\times3=6{,}120$ selected attempts. The same bug cohort and attempt
selection apply to accuracy, cost, variability, stratified results, and both
kinds of CI.

All primary metrics are recomputed from these selected attempts before
aggregation. The released selection manifest records selected run IDs,
skipped attempts, extra unused valid attempts, the fixed bug cohort, and source
hashes. The portable score snapshot retains the selected per-attempt scores
and the common bug IDs, allowing both kinds of CI to be recomputed without
access to full trajectories. Exploratory subset and separate trace analyses
state their own cohorts.

\section{Failure-Mode Trace Details}
\label{app:failure-mode-details}

This appendix gives the operational details behind the failure-mode discussion
in \Cref{sec:evaluation}. The main category breakdown measures late
discovery at the outcome level: in cross-component opaque bugs, the report gives
an initial stack-trace anchor but does not name the causal code, and 12 of the 17
configurations stay below 40\% block coverage. The diagnostics below quantify the same failure
modes using post hoc measurements over execution traces; they are not part of
benchmark scoring.

\subsection{Trace Artifacts}

Each completed run stores the final structured diagnosis, run statistics, and a
chronological trajectory:

\begin{Verbatim}[frame=single]
run-XX/
  diagnosis.json         # final structured diagnosis emitted by the harness
  diagnosis_scores.json  # overlap scores against human labels
  session_stats.json     # cost, token, timing, and call-count summary
  trajectory.json        # model messages plus tool calls and tool outputs
\end{Verbatim}

The run directories of all 6,120 scored attempts are released at \tracesurl.
\texttt{paper/reproduce.py} in \benchrepo rescores them with the released
evaluator and rebuilds \Cref{tab:eval}, \Cref{fig:bench-characterization},
and \Cref{fig:eval-evidence-block-coverage}.

The trace analyses below read \texttt{trajectory.json} after evaluation has
finished. They do not change scoring. We use the traces to inspect which files,
logs, and symbols each run searched before producing its final diagnosis.

\subsection{Late-Discovery Proxy}

For runs with parseable execution traces, we define \emph{late recovery} as a
run where none of the final diagnosis files appears among the first two
inspected source files. This is a conservative proxy for recovering from an
early search miss. It does not prove that the agent reasoned by backtracking,
but it identifies cases where the final answer was not already present in the
first local source-code scope.

\begin{table}[tbp]
  \centering
  \small
  \caption{Late-recovery proxy over parseable Mini-SWE-Agent and OpenCode traces.}
  \label{tab:appendix-late-recovery}
  \begin{tabular}{lrrr}
    \toprule
    Harness & Runs with early scope & Late recoveries & Rate \\
    \midrule
    Mini-SWE-Agent & 21 & 7 & 33.3\% \\
    OpenCode & 103 & 19 & 18.4\% \\
    Combined & 124 & 26 & 21.0\% \\
    \bottomrule
  \end{tabular}
\end{table}

Among the 80 successful runs in this trace subset, only 9 are late recoveries
(11.25\%). Current harnesses are therefore strongly shaped by early search
scope: they can follow evidence inside an initial region, but successful
diagnoses rarely arise after the correct file was absent from the first two
inspected source files.

\subsection{Tool-Call Nondeterminism Scan}

Tool-call nondeterminism is a concrete driver of diagnosis nondeterminism. We
ran a lightweight scan over Mini-SWE-Agent trajectories on the 120-bug main
evaluation set. For each bug and model, we compared the first five shell
commands across the three completed runs. We also extracted the first source
file path mentioned in a file-reading or code-search command. This measures
early search divergence, not correctness.

\begin{table}[tbp]
  \centering
  \small
  \caption{Early tool-use divergence across three Mini-SWE-Agent attempts per bug.}
  \label{tab:appendix-tool-nondeterminism}
  \begin{tabular}{lrr}
    \toprule
    Model & Bugs with different first-five commands & Bugs with different first source file \\
    \midrule
    GPT-5.4 & 120/120 & 29/120 \\
    GPT-5.3-Codex & 120/120 & 27/120 \\
    GPT-5.2-Codex & 120/120 & 32/120 \\
    \bottomrule
  \end{tabular}
\end{table}

The first-five-command measure is intentionally coarse and should not be read as
an error rate: harmless differences such as command flags or ordering are enough
to count as divergence. The first-source-file measure is more diagnosis-relevant.
It shows that a non-trivial fraction of repeated attempts begin reading
different code regions. On weak-evidence bugs, those early differences can
determine which local explanation the run later tries to justify. This motivates
adaptive tooling that augments the agent's context automatically when the early
search scope is incomplete, rather than relying on a single sequence of manual
shell commands to discover the right evidence.

\subsection{Representative Trace Excerpts}

\Cref{tab:appendix-trace-excerpts} shows compact excerpts from raw trajectories.
The table paraphrases long shell commands and tool outputs. Each row compares
two attempts from the same harness and model on the same bug; it is not a
cross-model comparison. The point of the excerpts is to make the branch point
visible: small changes in search terms, code windows, and follow-up files can
lock the same model into different local explanations.

\begin{table}[tbp]
  \centering
  \small
  \caption{Representative same-model trajectory contrasts behind the
  failure-mode analysis.}
  \label{tab:appendix-trace-excerpts}
  \begin{tabular}{p{0.18\columnwidth}p{0.36\columnwidth}p{0.36\columnwidth}}
    \toprule
    Case & Lower-scoring tool path & Higher-scoring tool path \\
    \midrule
    ZooKeeper-3769,
    Mini-SWE HY3 &
    Greps the node logs for election states and notifications, then follows the
    connection-management branch into \texttt{QuorumCnxManager} and
    \texttt{haveDelivered}. The final diagnosis blames connection delivery and
    has 0 block coverage. &
    Greps the same logs for \texttt{ERROR}, \texttt{Exception}, and
    \texttt{WorkerReceiver}, then opens the receive loop in
    \texttt{FastLeaderElection.java}. The final diagnosis covers
    \texttt{WorkerReceiver.run}. \\
    \addlinespace
    ZooKeeper-2845,
    Mini-SWE GPT-5.3-Codex &
    Searches \texttt{retain}, \texttt{ZKDatabase}, \texttt{QuorumPeer},
    \texttt{Learner}, and \texttt{DIFF}/\texttt{TRUNC}. The run inspects
    database loading and sync code, then diagnoses quorum synchronization; it
    misses \texttt{ZooKeeperServer.shutdown}. &
    Starts from similar retain-database searches, but adds
    \texttt{clear}/\texttt{shutdown} queries and opens the
    \texttt{ZooKeeperServer} shutdown window. The final diagnosis covers the
    lifecycle block that retains stale database state. \\
    \bottomrule
  \end{tabular}
\end{table}

These examples show how tool-call nondeterminism becomes diagnosis
nondeterminism. In ZooKeeper-3769, changing the log query from election-state
terms to exception and receiver-thread terms changes the next inspected file and
the final causal story. In ZooKeeper-2845, both attempts start with the same
retain-database vocabulary, but one keeps deepening the sync hypothesis while
the other asks which shutdown path preserves the stale state.

\subsection{Attempt-Level Headroom}

The repeated-run setup also gives a simple upper-bound check on whether agents
could do better with more reliable execution and self-correction. For each bug,
we compare the reported mean block coverage across three attempts with an oracle
that takes the best attempt for that same bug. This is not a deployable metric,
but it shows whether failures are entirely fixed by model capability or whether
execution consistency leaves measurable headroom.

\begin{table}[tbp]
  \centering
  \small
  \caption{Best-attempt headroom across the three selected Mini-SWE-Agent attempts per bug on the 120-bug common cohort.}
  \label{tab:appendix-best-attempt-headroom}
	\begin{tabular}{lrrrr}
	  \toprule
	  Model & Mean block & Best-attempt block & Gain & Bugs with better attempt \\
	  \midrule
	  GPT-6 Astra & 72.6\% & 76.5\% & 3.8 & 16/120 \\
	  Claude Opus 5.5 & 71.8\% & 78.4\% & 6.5 & 21/120 \\
	  GPT-5.5 & 70.0\% & 76.7\% & 6.7 & 25/120 \\
	  GPT-5.4 & 66.6\% & 73.7\% & 7.1 & 25/120 \\
	  GPT-5.3-Codex & 63.2\% & 69.8\% & 6.5 & 26/120 \\
	  GPT-5.2-Codex & 59.5\% & 64.3\% & 4.8 & 21/120 \\
    \bottomrule
  \end{tabular}
\end{table}

For GPT-5.5, 16 bugs are fully covered in at least one attempt but not across
all three attempts. This means some failures are not caused by an absolute
inability to identify the right source region. The same model and harness can
sometimes find the right blocks, but the execution path is not reliable enough
to do so consistently.

\subsection{ZooKeeper-2845 Trace Pattern}

ZooKeeper-2845 is an example of early scope lock-in. The report describes
retained database state during leader election, a transaction log ahead of a
snapshot, and an inconsistent quorum after restart. This language makes quorum
and sync code a plausible first target. Several traces therefore begin by
searching or reading under \texttt{src/java/main/org/apache/zookeeper/server/quorum}.

The gold block, however, is in:

\begin{Verbatim}[frame=single]
src/java/main/org/apache/zookeeper/server/ZooKeeperServer.java
function: ZooKeeperServer.shutdown
lines: 555--558
\end{Verbatim}

Runs that stay inside quorum code can build a coherent explanation around leader
election or synchronization while missing the lifecycle block that controls
database retention. The agent is not failing to find any relevant code; it is
stopping at the most obvious symptom-side location before looking for the
earlier state transition that made the symptom possible. When the report does
not directly name the exact causal code, the initial issue narrative can steer
the agent toward a nearby but incomplete diagnosis.

\subsection{Implication for Future Harnesses}

These observations suggest three concrete harness requirements. First, the
harness should provide adaptive context construction: when an early scope is
under-supported, tooling should surface adjacent files, call paths, dataflow
edges, related symbols, and alternative subsystems. Second, the harness should
preserve hypotheses as separate state, so evidence collected for one branch does
not become the default context for all later reasoning. Third, the harness
should add correctness checks that ask whether the proposed source region
explains the whole causal chain, rather than merely matching a symptom, stack
frame, or local code path.

\section{Exploratory Subset Experiments}
\label{app:exploratory-subset-experiments}

We also ran exploratory experiments with DeepSeek and Qwen models under
Mini-SWE-Agent and OpenCode. These runs are useful for model-selection context
but are not leaderboard-comparable: each model completed a different subset of
bugs. To make the comparison fairer, we compare each model only against Mini-SWE
GPT-5.2-Codex on the exact same completed bug subset.

\begin{table*}[t]
  \centering
  \scriptsize
  \setlength{\tabcolsep}{2.9pt}
  \caption{Exploratory subset experiments. Each model is compared against Mini-SWE GPT-5.2-Codex on exactly the subset of bugs completed by that model. Cov., prec., and std. abbreviate coverage, precision, and standard deviation; all are percentages. Mini-SWE denotes Mini-SWE-Agent, and subset reports completed bugs / represented systems. Parenthesized values are paired deltas from GPT-5.2-Codex on the same subset. Flaky bugs are counts, and flaky delta is the model count minus the GPT-5.2-Codex count. Token usage is average tokens per run, with multiplier relative to GPT-5.2 on the same bugs.}
  \label{tab:exploratory-subset-experiments}
  \begin{tabular}{@{}lllrrrrrrr@{}}
    \toprule
    Model & Harness & Subset & Block cov. & Block prec. & Mention cov. & Cov. std. & Flaky bugs & Flaky delta & Tokens/run \\
    \midrule
    DeepSeek V4 Pro & Mini-SWE & 67/10 & 64.8 (+5.6) & 41.6 (-3.5) & 74.0 (+11.8) & 8.8 (+4.7) & 17 & +8 & 1.4M (2.5$\times$) \\
    Qwen3-Coder-Next & Mini-SWE & 55/8 & 59.9 (+1.0) & 47.4 (+3.2) & 66.5 (+4.7) & 17.1 (+12.8) & 25 & +18 & 4.4M (7.7$\times$) \\
    Qwen3.6 Max & Mini-SWE & 35/7 & 61.8 (+6.4) & 48.7 (+10.5) & 70.2 (+11.8) & 14.2 (+9.6) & 15 & +10 & 1.1M (1.9$\times$) \\
    Qwen3.6 Max & OpenCode & 63/10 & 53.6 (-4.0) & 46.7 (+4.2) & 63.8 (+3.0) & 8.2 (+3.5) & 15 & +6 & 0.6M (0.9$\times$) \\
    \bottomrule
  \end{tabular}
\end{table*}

\begin{figure*}[t]
  \centering
  \includegraphics[width=\textwidth]{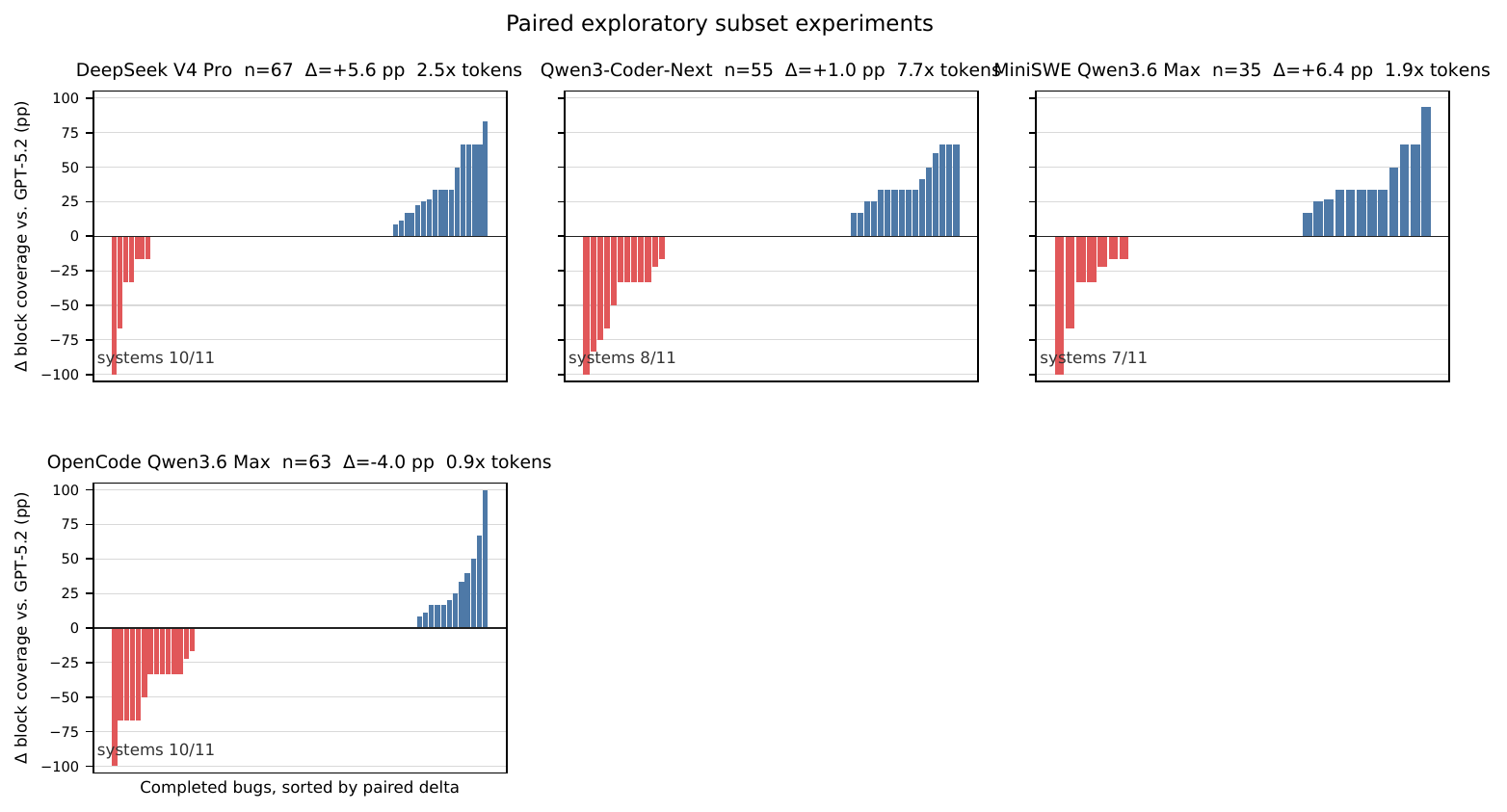}
  \caption{Paired per-bug block-coverage deltas for exploratory subset
  experiments. Each bar is one bug completed by the model, compared against
  Mini-SWE GPT-5.2-Codex on the same bug. Positive bars mean the exploratory
  model outperformed the baseline on that bug. These runs are not in the main
  leaderboard because the completed subsets are non-uniform.}
  \label{fig:subset-experiment-deltas}
\end{figure*}

\begin{figure*}[t]
  \centering
  \includegraphics[width=\textwidth]{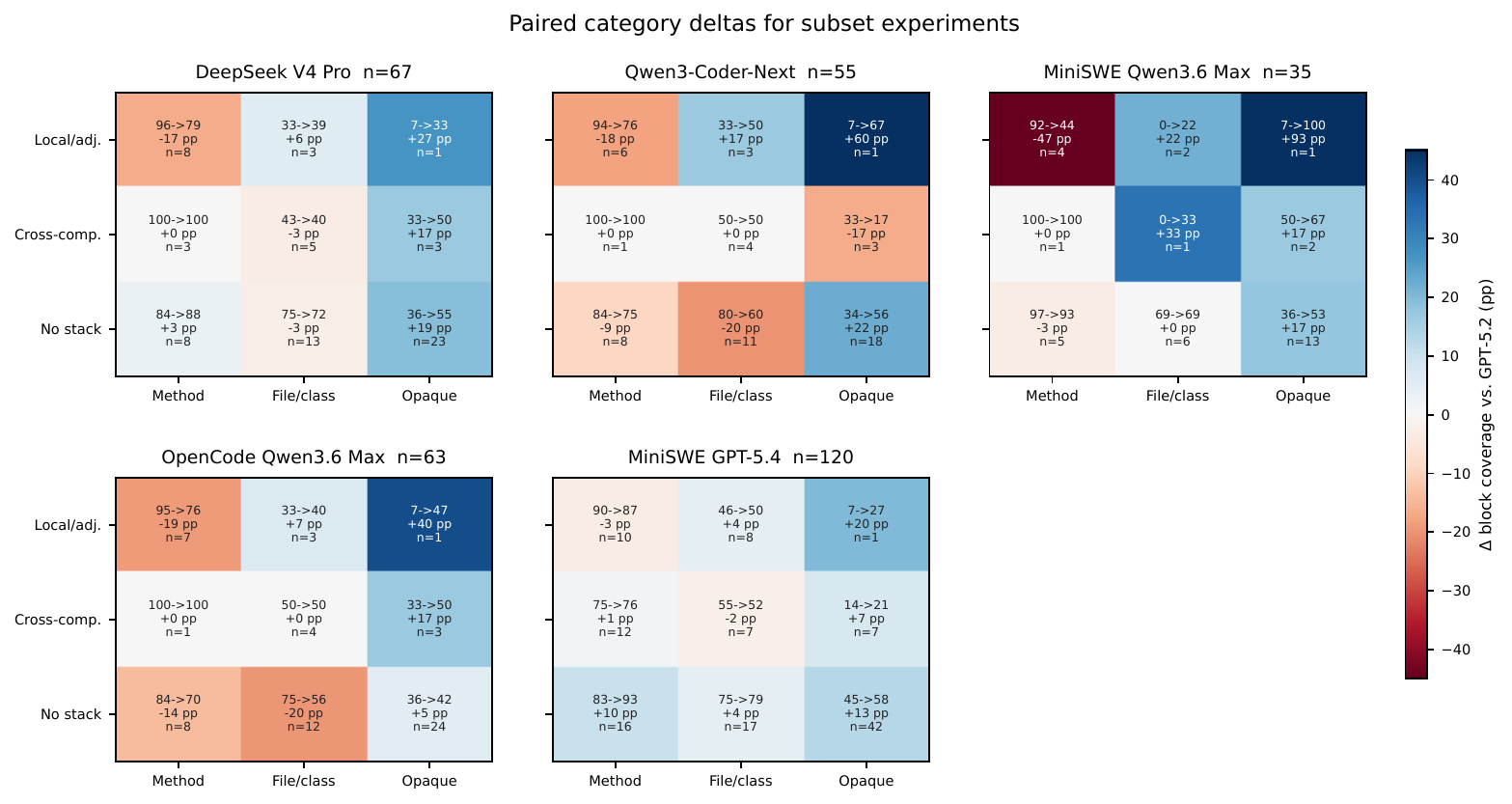}
  \caption{Paired block-coverage deltas by evidence category for subset experiments.
  Each cell compares a model with Mini-SWE GPT-5.2-Codex on the same completed
  bug subset and category. The GPT-5.4 panel uses the 120-bug shared
  evaluation set; the other panels are exploratory subset experiments. Cell text
  shows absolute GPT-5.2$\rightarrow$model block coverage, the paired delta in
  percentage points, and the number of bugs. Counts are small in some cells, so
  the plot is descriptive rather than a leaderboard comparison.}
  \label{fig:subset-category-deltas}
\end{figure*}

\begin{figure*}[t]
  \centering
  \includegraphics[width=\textwidth]{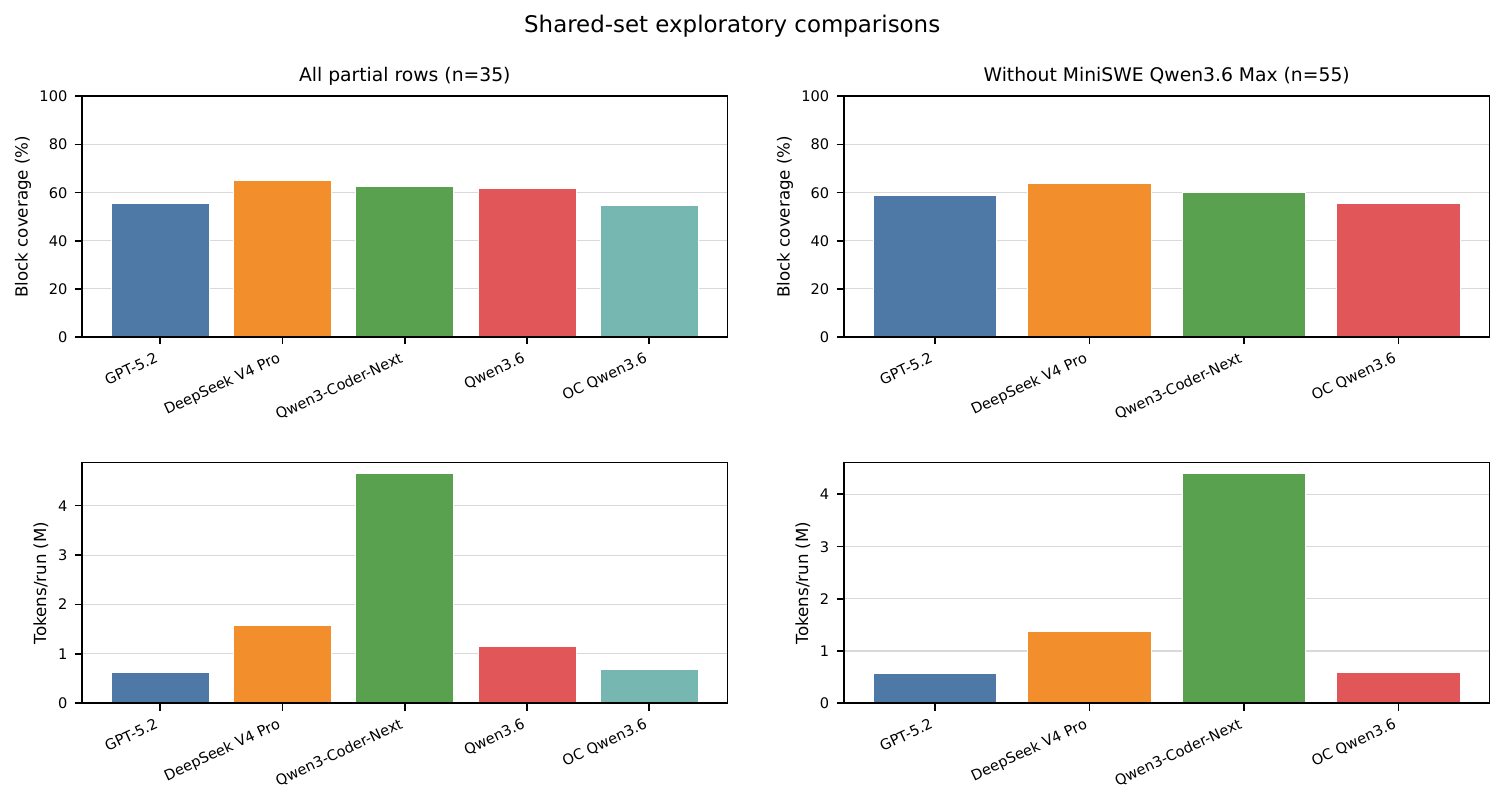}
  \caption{Shared-set exploratory comparisons. The left column uses the 35 bugs
  completed by all partial exploratory rows and Mini-SWE GPT-5.2-Codex. The right
  column removes the most constraining Mini-SWE Qwen3.6 Max run, yielding 55
  shared bugs. The bottom row shows that several newer models require
  substantially more tokens even when evaluated on the same bug set.}
  \label{fig:shared-subset-comparison}
\end{figure*}

\Cref{tab:exploratory-subset-experiments} shows that several exploratory models match
or exceed the GPT-5.2-Codex baseline on their own completed subsets, but often
with much higher token usage. \Cref{fig:subset-experiment-deltas} shows the
paired per-bug deltas, while \Cref{fig:subset-category-deltas} shows the
same paired comparison broken down by evidence category. Finally,
\Cref{fig:shared-subset-comparison} compares models on common shared subsets:
the 35 bugs completed by all partial exploratory rows, and a 55-bug set after
removing the most constraining Mini-SWE Qwen3.6 Max run. These results support
the main-paper decision to use a common 120-bug cohort with three valid scored attempts per configuration in the primary
leaderboard while treating non-uniform subset experiments as exploratory evidence
rather than primary benchmark results.

\clearpage
\section{Evaluator Validity and Contamination Analyses}
\label{app:validity-evidence}

These studies test whether the localization metrics capture useful diagnosis
information and whether public-issue exposure can explain the results.

\subsection{Evaluator Validation}
\label{app:evaluator-validation}

\paragraph{Score-blinded audit.}
We audit 20 GPT-5.4 diagnoses selected for zero block coverage, with the scores
hidden during qualitative assessment. The assessment distinguishes whether a
diagnosis reconstructs the failure mechanism from whether it places
responsibility at the annotated code location. \Cref{tab:diagnosis-audit} gives
the complete outcome breakdown. Four diagnoses reconstruct the causal chain
but identify a different plausible intervention point. For example, one places
responsibility for a nil-pointer failure in a specific \texttt{Op.check}
implementation, while the accepted localization is in the broader
\texttt{processRequest} path. The remaining 16 diagnoses have an incorrect
hypothesis or an incomplete causal explanation.

\begin{table}[htbp]
  \centering\small
  \caption{Score-blinded audit of 20 zero-block-coverage GPT-5.4 diagnoses.}
  \label{tab:diagnosis-audit}
  \begin{tabular}{@{}lr@{}}\toprule
  Outcome & Diagnoses \\
  \midrule
  Complete causal chain, different plausible intervention point & 4 \\
  Incorrect hypothesis & 9 \\
  Missing the decisive implementation link & 2 \\
  Superficial or incomplete causal account & 5 \\
  \midrule
  Total & 20 \\
  \bottomrule\end{tabular}
\end{table}

Plausible alternative localizations therefore do not explain most zero-score
outputs: 16/20 are incorrect or causally incomplete. Block coverage measures
recovery of the annotated causal implementation rather than every plausible
intervention or repair point, and the four alternative diagnoses show where
this distinction matters. Because the audit samples only zero-scoring outputs,
these proportions describe the sample rather than a benchmark-wide
false-negative rate.

\paragraph{Benchmark-wide false-negative estimate.}
We estimate that rate from a stratified random sample of the 3,134 attempts
with incomplete block coverage across all 17 configurations: 92 attempts, 60
with zero and 32 with partial coverage, roughly in proportion to the two strata.
The sample contains 178 missed gold blocks. Two AI reviewers judge each missed block
independently: Claude Opus 5.5 and GPT-6 Astra (run through Claude Code and Codex
CLI), both at high reasoning effort, with read-only access to the pre-fix source
and no network access. Each receives the issue report, the gold blocks, the
historical fix as ground truth, and the diagnosis, without its score or the model
that produced it. A missed block is a false negative when the diagnosis names
the same code at shifted coordinates, or a different location that is itself
the defect: changing it alone would prevent the reported failure, and it is not
a caller, a symptom site, or a defensive check. We count a false negative only
when both reviewers agree.

The reviewers agree on 87\% of missed blocks (Cohen's $\kappa = 0.36$).
Weighting each stratum by its number of missed blocks, 4.5\% of missed gold
blocks are false negatives, and counting them would raise block coverage by 2.3
points (\Cref{tab:fn-estimate}). Of the eight false-negative blocks, one names
the gold code at different coordinates, and the other seven are alternative
locations. Either reviewer alone yields a higher rate, so requiring agreement
gives the conservative estimate.

\begin{table}[htbp]
  \centering\small
  \caption{Benchmark-wide evaluator false negatives, estimated from 92 sampled attempts with incomplete block coverage (178 missed gold blocks). Brackets are 95\% stratified bootstrap intervals; the first row is the reported estimate.}
  \label{tab:fn-estimate}
  \begin{tabular}{@{}lcc@{}}\toprule
  Counting rule & False negatives among missed blocks & Block coverage gain (points) \\
  \midrule
  Both reviewers agree & 4.5\% [1.6, 8.3] & 2.3 [0.7, 4.3] \\
  Claude Opus 5.5 alone & 5.6\% [2.3, 9.8] & 3.1 [1.2, 5.5] \\
  GPT-6 Astra alone & 16.3\% [10.0, 23.9] & 7.2 [4.2, 10.6] \\
  \bottomrule\end{tabular}
\end{table}

\paragraph{Diagnosis-assisted repair: design.}
We select 20 bugs from the GPT-5.4 diagnosis evaluation: a ten-bug high-coverage
set with an available 100\%-coverage diagnosis and a ten-bug low-coverage set
with an available 0\%-coverage diagnosis. Six sampled bugs have both high- and
low-coverage diagnoses from the same model and harness. These mixed cases
support a within-bug comparison with three conditions: no supplied diagnosis,
a low-coverage diagnosis, and a high-coverage diagnosis. For the other cases,
we compare no diagnosis with the available diagnosis condition. Each condition
has three repair attempts per bug. In the six-bug comparison, the fixing model,
repository, harness, tools, and budget are held fixed; the supplied diagnosis
changes.

One author inspects each generated patch against the original issue resolution
and PR. \emph{Gold-file touch} indicates that the patch modifies a file where
the reference repair should land; \emph{gold-block touch} indicates that it
modifies the annotated causal region. \emph{Patch produced} reports whether a
patch was generated.

\begin{table}[htbp]
  \centering\small
  \caption{Diagnosis-assisted repair on the two ten-bug sets (30 attempts per
  row). Patch production and touch rates are percentages; cost is mean API
  cost per repair attempt in USD.}
  \label{tab:repair-pilot-sets}
  \setlength{\tabcolsep}{4pt}
  \begin{tabular}{@{}llrrrrr@{}}\toprule
  Set & Diagnosis & Patch & File touch & Block touch & Calls/run & Cost \\
  \midrule
  High & None & 100.0 & 86.7 & 56.7 & 34.6 & \$1.44 \\
  High & High coverage & 100.0 & 90.0 & 60.0 & 30.1 & \$1.02 \\
  Low & None & 93.3 & 26.7 & 6.7 & 40.4 & \$2.00 \\
  Low & Low coverage & 100.0 & 20.0 & 0.0 & 40.6 & \$1.74 \\
  \bottomrule\end{tabular}
\end{table}

\begin{table}[htbp]
  \centering\small
  \caption{Within-bug repair comparison on six mixed cases, with three attempts
  per condition per bug. Touch rates are percentages with raw counts.}
  \label{tab:repair-pilot-mixed}
  \setlength{\tabcolsep}{4pt}
  \begin{tabular}{@{}lrrrrr@{}}\toprule
  Diagnosis & Attempts & File touch & Block touch & Calls/run & Cost \\
  \midrule
  None & 18 & 44.4 (8/18) & 11.1 (2/18) & 40.9 & \$2.02 \\
  Low coverage & 18 & 33.3 (6/18) & 0.0 (0/18) & 37.9 & \$1.54 \\
  High coverage & 18 & 61.1 (11/18) & 50.0 (9/18) & 31.6 & \$1.34 \\
  \bottomrule\end{tabular}
\end{table}

\paragraph{Repair results.}
On the six mixed cases, replacing a low-coverage diagnosis with a high-coverage
one raises gold-file touch by 27.8 percentage points and gold-block touch by
50.0 points (\Cref{tab:repair-pilot-mixed}). Relative to no diagnosis, the
high-coverage diagnosis raises gold-block touch by 38.9 points while reducing
mean model calls by 22.7\% and API cost by 33.7\%. The ten-bug sets show a
similar pattern (\Cref{tab:repair-pilot-sets}). On the high-coverage set, the
unaided agent already touches a gold block in 56.7\% of attempts, so the
diagnosis mainly reduces search effort: gold-block touch rises to 60.0\%,
while cost falls from \$1.44 to \$1.02. On the low-coverage set, no diagnosis
is better than a bad one: gold-block touch falls from 6.7\% to 0\%.

This is a small pilot rather than a precise estimate of end-to-end patch
correctness. Nevertheless, the controlled within-bug comparison provides
evidence that diagnoses with higher block coverage guide downstream repair
toward the causal code more reliably and efficiently.

\subsection{Contamination Analyses}
\label{app:contamination-evidence}

\paragraph{Direct answer retrieval and the leakage guard.}
Our reward-hacking check targets a concrete shortcut: retrieving the historical
fix instead of diagnosing the failure from the supplied incident evidence and
pre-fix code. Early experiments ran without the leakage guard. Subsequent trace
inspection revealed agents searching for and retrieving future fixing commits.
Workspace paths also exposed the original issue ID, giving agents a lookup key.
We therefore removed commits after the task snapshot and anonymized workspace
directory names.

The guard scans visible reasoning, tool inputs, and tool outputs for the
original issue ID that should have been hidden from the agent. Matches include
the full issue key or its numeric component, such as
\texttt{git log | grep ZooKeeper-2845} and \texttt{git log | grep 2845}.
We apply this check to recorded trajectories, including the early experiments,
and discard flagged attempts, rerunning until every bug has three clean
attempts. Residual hits also prompt an audit of
the prepared workspace to remove exposed identifiers or fix information.
\Cref{tab:leakage-guard-results} reports affected trajectories separately from
matching events in each channel; a single trajectory may contain many matches.
The Codex CLI and OpenCode prompt templates included the issue identifier.
In their 1,440 selected attempts, 34 searched the workspace with it, none
retrieved fix information, and these attempts scored lower than the rest of
their configuration.

\begin{table}[htbp]
  \centering\small
  \caption{Leakage-guard results before and after removing future commits.
  The reporting denominator is 360 attempts ($120$ bugs $\times$ $3$ attempts).
  Parentheses give the percentage flagged; the last three columns count
  matching events.}
  \label{tab:leakage-guard-results}
  \setlength{\tabcolsep}{4pt}
  \begin{tabular}{@{}llrrrr@{}}\toprule
  Phase & Model & Flagged (\%) & Reasoning & Tool output & Tool input \\
  \midrule
  Before & GPT-5.4 & 102 (28.3) & 0 & 1,007 & 16 \\
  After & GPT-5.4 & 4 (1.1) & 0 & 22 & 0 \\
  After & GPT-5.5 & 4 (1.1) & 0 & 22 & 2 \\
  After & GPT-6 Astra & 3 (0.8) & 0 & 10 & 2 \\
  After & Claude Opus 5.5 & 0 (0.0) & 0 & 0 & 0 \\
  Before & GPT-5.3-Codex & 8 (2.2) & 0 & 88 & 0 \\
  \bottomrule\end{tabular}
\end{table}

Removing future commits substantially reduces detected leakage: GPT-5.4's
flagged attempts fall from 102/360 (28.3\%) to 4/360 (1.1\%). The remaining
hits show why trace checks are still needed after workspace cleanup.

\paragraph{Implicit memorization.}
Identifier matching detects explicit traces of answer retrieval, but a model
could recall a bug without mentioning its ID. Long-public bugs can still be
difficult: in ZooKeeper-2845, failed runs focus on the quorum subsystem suggested
by the report and miss the causal code in \texttt{ZooKeeperServer.java}
(\Cref{app:failure-mode-details}). We examine this question more broadly through
issue-age correlations and closed-book diagnosis probes.

\paragraph{Issue age and public discussion.}
Across all 120 issues, GPT-5.4 block coverage has Pearson $r=-0.027$ and
Spearman $\rho=-0.041$ with issue age, and $r=0.008$ and $\rho=0.011$
with the current issue-comment count. Across the 13 Mini-SWE configurations,
issue-age correlations range from $-0.112$ to $0.062$ (Pearson) and from
$-0.141$ to $0.053$ (Spearman; \Cref{tab:issue-age-correlations}). Older or more
discussed issues are not systematically easier to diagnose. GPT-5.4 coverage is
negatively associated with time to resolution ($r=-0.239$ after removing
per-system means and controlling for issue age), which is more consistent with
difficult bugs taking longer to resolve than with greater public exposure making
them easier.

\begin{table}[htbp]
  \centering\small
  \caption{Correlations between issue age and bug-level block coverage for the
  13 Mini-SWE configurations. $r^2$ values are percentages.}
  \label{tab:issue-age-correlations}
  \begin{tabular}{@{}lrrr@{}}\toprule
  Model & Pearson $r$ & Spearman $\rho$ & Pearson $r^2$ (\%) \\
  \midrule
  GPT-6 Astra & $0.004$ & $-0.001$ & 0.00 \\
  Claude Opus 5.5 & $-0.029$ & $-0.034$ & 0.08 \\
  GPT-5.5 & $-0.051$ & $-0.042$ & 0.26 \\
  GPT-5.4 & $-0.027$ & $-0.041$ & 0.07 \\
  GPT-5.3-Codex & $-0.044$ & $-0.070$ & 0.19 \\
  MiMo v2.5 Pro & $0.062$ & $0.053$ & 0.39 \\
  Kimi K2.6 & $-0.022$ & $-0.065$ & 0.05 \\
  MiMo v2.5 & $-0.030$ & $-0.048$ & 0.09 \\
  GPT-5.2-Codex & $-0.059$ & $-0.097$ & 0.35 \\
  DeepSeek V4 Flash & $-0.078$ & $-0.121$ & 0.60 \\
  HY3 & $-0.087$ & $-0.112$ & 0.76 \\
  Gemini 3.5 Flash & $0.030$ & $0.002$ & 0.09 \\
  o4-mini & $-0.112$ & $-0.141$ & 1.25 \\
  \bottomrule\end{tabular}
\end{table}

\paragraph{Closed-book probes: design.}
We evaluate GPT-5.4 without repository access, tools, Git history, or internet.
Of the 120 benchmark reports, we exclude 54 that directly contain an exact
gold filename or function name, leaving 66 bugs. Three attempts are averaged
within each bug before averaging across bugs. We compare three prompt variants:
the original report with its issue ID (identifiable), the report with its issue
ID redacted (anonymous), and only the project and issue ID without the report
(ID only). The repository-assisted reference in
\Cref{tab:closed-book-results} uses the same 66 bugs. Without source access, the
model does not emit source line ranges, so closed-book line and block coverage
are zero; we compare variants on file and function coverage.

\begin{table}[htbp]
  \centering\small
  \caption{GPT-5.4 contamination probes on 66 bugs (198 attempts per
  closed-book variant). All entries are mean coverage percentages.}
  \label{tab:closed-book-results}
  \begin{tabular}{@{}lrrrr@{}}\toprule
  Condition & File & Function & Line & Block \\
  \midrule
  Repository-assisted & 76.0 & 46.7 & 46.9 & 58.3 \\
  Closed book: identifiable report & 43.9 & 17.1 & 0.0 & 0.0 \\
  Closed book: anonymous report & 42.7 & 21.1 & 0.0 & 0.0 \\
  Closed book: issue ID only & 0.0 & 0.0 & 0.0 & 0.0 \\
  \bottomrule\end{tabular}
\end{table}

\paragraph{Closed-book results and interpretation.}
Repository-assisted coverage exceeds every closed-book variant. Providing the
issue ID does not improve function coverage in these probes: identifiable reports
yield 17.1\%, compared with 21.1\% for anonymous reports. ID-only prompts
yield zero coverage at all four granularities. A separate report-to-ID probe
recovers 0/360 exact target identifiers. In contrast, repository-assisted
function coverage reaches 46.7\% and block coverage reaches 58.3\%. The gap
supports codebase exploration as an important part of diagnosis: report text
and issue identity alone do not recover the same causal-code detail.

Nonzero file/function coverage without repository access can reflect knowledge
of project structure or clues in reports, including class and subsystem names
that remain after exact gold filenames and function names are removed. For
example, references to \texttt{LocalHealthMonitor}, \texttt{InitProducerId}, or
\texttt{QuorumCnxManager.Listener} constrain plausible code regions.

Together, the trace audit identifies and removes direct answer-retrieval
shortcuts, while the age analysis and closed-book probes test whether public
issue exposure readily translates into accurate diagnosis. The results support
the importance of exploring the repository to identify causal code. They do not
rule out all pretraining exposure or implicit memorization.

\clearpage
\section*{NeurIPS Paper Checklist}

\begin{enumerate}

\item {\bf Claims}
    \item[] Question: Do the main claims made in the abstract and introduction accurately reflect the paper's contributions and scope?
    \item[] Answer: \answerYes{} %
    \item[] Justification: The abstract and \Cref{sec:intro} state the paper's scope: a benchmark for distributed-systems diagnosis, block-level root-cause labels, and an empirical evaluation of current agents.
    \item[] Guidelines:
    \begin{itemize}
        \item The answer \answerNA{} means that the abstract and introduction do not include the claims made in the paper.
        \item The abstract and/or introduction should clearly state the claims made, including the contributions made in the paper and important assumptions and limitations. A \answerNo{} or \answerNA{} answer to this question will not be perceived well by the reviewers. 
        \item The claims made should match theoretical and experimental results, and reflect how much the results can be expected to generalize to other settings. 
        \item It is fine to include aspirational goals as motivation as long as it is clear that these goals are not attained by the paper. 
    \end{itemize}

\item {\bf Limitations}
    \item[] Question: Does the paper discuss the limitations of the work performed by the authors?
    \item[] Answer: \answerYes{} %
    \item[] Justification: \Cref{app:benchmark-validity,app:eval-uncertainty}
    discuss benchmark selection, label sensitivity, small category samples,
    and the assumptions behind bug-level resampling.
    \Cref{app:evaluator-validation,app:contamination-evidence} describe the scope
    of the human audit and repair pilot, identifier-based leakage detection,
    and the limits of the memorization probes.
    \item[] Guidelines:
    \begin{itemize}
        \item The answer \answerNA{} means that the paper has no limitation while the answer \answerNo{} means that the paper has limitations, but those are not discussed in the paper. 
        \item The authors are encouraged to create a separate ``Limitations'' section in their paper.
        \item The paper should point out any strong assumptions and how robust the results are to violations of these assumptions (e.g., independence assumptions, noiseless settings, model well-specification, asymptotic approximations only holding locally). The authors should reflect on how these assumptions might be violated in practice and what the implications would be.
        \item The authors should reflect on the scope of the claims made, e.g., if the approach was only tested on a few datasets or with a few runs. In general, empirical results often depend on implicit assumptions, which should be articulated.
        \item The authors should reflect on the factors that influence the performance of the approach. For example, a facial recognition algorithm may perform poorly when image resolution is low or images are taken in low lighting. Or a speech-to-text system might not be used reliably to provide closed captions for online lectures because it fails to handle technical jargon.
        \item The authors should discuss the computational efficiency of the proposed algorithms and how they scale with dataset size.
        \item If applicable, the authors should discuss possible limitations of their approach to address problems of privacy and fairness.
        \item While the authors might fear that complete honesty about limitations might be used by reviewers as grounds for rejection, a worse outcome might be that reviewers discover limitations that aren't acknowledged in the paper. The authors should use their best judgment and recognize that individual actions in favor of transparency play an important role in developing norms that preserve the integrity of the community. Reviewers will be specifically instructed to not penalize honesty concerning limitations.
    \end{itemize}

\item {\bf Theory assumptions and proofs}
    \item[] Question: For each theoretical result, does the paper provide the full set of assumptions and a complete (and correct) proof?
    \item[] Answer: \answerNA{} %
    \item[] Justification: The paper does not present theoretical results or proofs.
    \item[] Guidelines:
    \begin{itemize}
        \item The answer \answerNA{} means that the paper does not include theoretical results. 
        \item All the theorems, formulas, and proofs in the paper should be numbered and cross-referenced.
        \item All assumptions should be clearly stated or referenced in the statement of any theorems.
        \item The proofs can either appear in the main paper or the supplemental material, but if they appear in the supplemental material, the authors are encouraged to provide a short proof sketch to provide intuition. 
        \item Inversely, any informal proof provided in the core of the paper should be complemented by formal proofs provided in appendix or supplemental material.
        \item Theorems and Lemmas that the proof relies upon should be properly referenced. 
    \end{itemize}

    \item {\bf Experimental result reproducibility}
    \item[] Question: Does the paper fully disclose all the information needed to reproduce the main experimental results of the paper to the extent that it affects the main claims and/or conclusions of the paper (regardless of whether the code and data are provided or not)?
    \item[] Answer: \answerYes{} %
    \item[] Justification: \Cref{sec:benchmark,sec:evaluation,app:prompts}
    specify task inputs, labels, model and harness versions, reasoning effort,
    prompt excerpts, and scoring metrics.
    \Cref{app:eval-attempt-counts,app:eval-uncertainty,app:eval-system-sensitivity}
    specify attempt selection, aggregation, bootstrap settings, paired
    comparisons, and system exclusions.
    \item[] Guidelines:
    \begin{itemize}
        \item The answer \answerNA{} means that the paper does not include experiments.
        \item If the paper includes experiments, a \answerNo{} answer to this question will not be perceived well by the reviewers: Making the paper reproducible is important, regardless of whether the code and data are provided or not.
        \item If the contribution is a dataset and\slash or model, the authors should describe the steps taken to make their results reproducible or verifiable. 
        \item Depending on the contribution, reproducibility can be accomplished in various ways. For example, if the contribution is a novel architecture, describing the architecture fully might suffice, or if the contribution is a specific model and empirical evaluation, it may be necessary to either make it possible for others to replicate the model with the same dataset, or provide access to the model. In general, releasing code and data is often one good way to accomplish this, but reproducibility can also be provided via detailed instructions for how to replicate the results, access to a hosted model (e.g., in the case of a large language model), releasing of a model checkpoint, or other means that are appropriate to the research performed.
        \item While NeurIPS does not require releasing code, the conference does require all submissions to provide some reasonable avenue for reproducibility, which may depend on the nature of the contribution. For example
        \begin{enumerate}
            \item If the contribution is primarily a new algorithm, the paper should make it clear how to reproduce that algorithm.
            \item If the contribution is primarily a new model architecture, the paper should describe the architecture clearly and fully.
            \item If the contribution is a new model (e.g., a large language model), then there should either be a way to access this model for reproducing the results or a way to reproduce the model (e.g., with an open-source dataset or instructions for how to construct the dataset).
            \item We recognize that reproducibility may be tricky in some cases, in which case authors are welcome to describe the particular way they provide for reproducibility. In the case of closed-source models, it may be that access to the model is limited in some way (e.g., to registered users), but it should be possible for other researchers to have some path to reproducing or verifying the results.
        \end{enumerate}
    \end{itemize}

\item {\bf Open access to data and code}
    \item[] Question: Does the paper provide open access to the data and code, with sufficient instructions to faithfully reproduce the main experimental results, as described in supplemental material?
    \item[] Answer: \answerYes{} %
    \item[] Justification: The benchmark, evaluator, and agent runner are released at \benchrepo, and the traces of all 6,120 scored attempts at \tracesurl, with a script that rebuilds \Cref{tab:eval} and \Cref{fig:bench-characterization,fig:eval-evidence-block-coverage} from them; \Cref{sec:benchmark} describes the released data format and \Cref{sec:evaluation} describes the evaluation protocol.
    \item[] Guidelines:
    \begin{itemize}
        \item The answer \answerNA{} means that the paper does not include experiments requiring code.
        \item Please see the NeurIPS code and data submission guidelines (\url{https://neurips.cc/public/guides/CodeSubmissionPolicy}) for more details.
        \item While we encourage the release of code and data, we understand that this might not be possible, so \answerNo{} is an acceptable answer. Papers cannot be rejected simply for not including code, unless this is central to the contribution (e.g., for a new open-source benchmark).
        \item The instructions should contain the exact command and environment needed to run to reproduce the results. See the NeurIPS code and data submission guidelines (\url{https://neurips.cc/public/guides/CodeSubmissionPolicy}) for more details.
        \item The authors should provide instructions on data access and preparation, including how to access the raw data, preprocessed data, intermediate data, and generated data, etc.
        \item The authors should provide scripts to reproduce all experimental results for the new proposed method and baselines. If only a subset of experiments are reproducible, they should state which ones are omitted from the script and why.
        \item At submission time, to preserve anonymity, the authors should release anonymized versions (if applicable).
        \item Providing as much information as possible in supplemental material (appended to the paper) is recommended, but including URLs to data and code is permitted.
    \end{itemize}

\item {\bf Experimental setting/details}
    \item[] Question: Does the paper specify all the training and test details (e.g., data splits, hyperparameters, how they were chosen, type of optimizer) necessary to understand the results?
    \item[] Answer: \answerYes{} %
    \item[] Justification: \Cref{sec:evaluation,app:prompts,app:eval-attempt-counts}
    give the evaluated configurations, task prompts, 120-bug cohort, and
    three-attempt selection protocol.
    \Cref{app:exploratory-subset-experiments,app:validity-evidence} specify the
    separate cohorts and conditions for subset evaluations, the repair pilot,
    and contamination probes.
    \item[] Guidelines:
    \begin{itemize}
        \item The answer \answerNA{} means that the paper does not include experiments.
        \item The experimental setting should be presented in the core of the paper to a level of detail that is necessary to appreciate the results and make sense of them.
        \item The full details can be provided either with the code, in appendix, or as supplemental material.
    \end{itemize}

\item {\bf Experiment statistical significance}
    \item[] Question: Does the paper report error bars suitably and correctly defined or other appropriate information about the statistical significance of the experiments?
    \item[] Answer: \answerYes{} %
    \item[] Justification: \Cref{app:eval-uncertainty} defines the 95\%
    bug-bootstrap intervals, including 10,000 resamples, seed 2026, the
    independence assumption, and paired resampling for model differences;
    \Cref{tab:eval-confidence-intervals,tab:eval-paired-confidence} report the results.
    \Cref{fig:eval-evidence-block-coverage,app:eval-system-sensitivity} give
    category counts and intervals and system sensitivity results, while
    \Cref{sec:evaluation} separately reports within-bug variability.
    \item[] Guidelines:
    \begin{itemize}
        \item The answer \answerNA{} means that the paper does not include experiments.
        \item The authors should answer \answerYes{} if the results are accompanied by error bars, confidence intervals, or statistical significance tests, at least for the experiments that support the main claims of the paper.
        \item The factors of variability that the error bars are capturing should be clearly stated (for example, train/test split, initialization, random drawing of some parameter, or overall run with given experimental conditions).
        \item The method for calculating the error bars should be explained (closed form formula, call to a library function, bootstrap, etc.)
        \item The assumptions made should be given (e.g., Normally distributed errors).
        \item It should be clear whether the error bar is the standard deviation or the standard error of the mean.
        \item It is OK to report 1-sigma error bars, but one should state it. The authors should preferably report a 2-sigma error bar than state that they have a 96\% CI, if the hypothesis of Normality of errors is not verified.
        \item For asymmetric distributions, the authors should be careful not to show in tables or figures symmetric error bars that would yield results that are out of range (e.g., negative error rates).
        \item If error bars are reported in tables or plots, the authors should explain in the text how they were calculated and reference the corresponding figures or tables in the text.
    \end{itemize}

\item {\bf Experiments compute resources}
    \item[] Question: For each experiment, does the paper provide sufficient information on the computer resources (type of compute workers, memory, time of execution) needed to reproduce the experiments?
    \item[] Answer: \answerYes{} %
    \item[] Justification: The dominant compute resource is hosted-model API
    inference; \Cref{sec:evaluation,tab:eval} identify the model/harness
    configurations and report API calls and tokens per diagnosis.
    \Cref{app:exploratory-subset-experiments,app:evaluator-validation} report
    token usage for subset evaluations and API costs for the repair pilot.
    \item[] Guidelines:
    \begin{itemize}
        \item The answer \answerNA{} means that the paper does not include experiments.
        \item The paper should indicate the type of compute workers CPU or GPU, internal cluster, or cloud provider, including relevant memory and storage.
        \item The paper should provide the amount of compute required for each of the individual experimental runs as well as estimate the total compute. 
        \item The paper should disclose whether the full research project required more compute than the experiments reported in the paper (e.g., preliminary or failed experiments that didn't make it into the paper). 
    \end{itemize}
    
\item {\bf Code of ethics}
    \item[] Question: Does the research conducted in the paper conform, in every respect, with the NeurIPS Code of Ethics \url{https://neurips.cc/public/EthicsGuidelines}?
    \item[] Answer: \answerYes{} %
    \item[] Justification: The work evaluates agents on public open-source
    issue reports and repositories; human involvement consists of author
    annotation, diagnosis assessment, and patch inspection
    (\Cref{app:benchmark-label-structure,app:evaluator-validation}).
    It does not recruit human participants or release a new model.
    \item[] Guidelines:
    \begin{itemize}
        \item The answer \answerNA{} means that the authors have not reviewed the NeurIPS Code of Ethics.
        \item If the authors answer \answerNo, they should explain the special circumstances that require a deviation from the Code of Ethics.
        \item The authors should make sure to preserve anonymity (e.g., if there is a special consideration due to laws or regulations in their jurisdiction).
    \end{itemize}

\item {\bf Broader impacts}
    \item[] Question: Does the paper discuss both potential positive societal impacts and negative societal impacts of the work performed?
    \item[] Answer: \answerYes{} %
    \item[] Justification: \Cref{sec:intro} motivates better diagnosis of costly
    production failures and discusses how repairs can suppress symptoms while
    leaving the root cause unresolved.
    \Cref{sec:evaluation,app:failure-mode-details} examine incorrect diagnoses
    and misleading validation signals and motivate stronger checks before
    acting on an agent's diagnosis.
    \item[] Guidelines:
    \begin{itemize}
        \item The answer \answerNA{} means that there is no societal impact of the work performed.
        \item If the authors answer \answerNA{} or \answerNo, they should explain why their work has no societal impact or why the paper does not address societal impact.
        \item Examples of negative societal impacts include potential malicious or unintended uses (e.g., disinformation, generating fake profiles, surveillance), fairness considerations (e.g., deployment of technologies that could make decisions that unfairly impact specific groups), privacy considerations, and security considerations.
        \item The conference expects that many papers will be foundational research and not tied to particular applications, let alone deployments. However, if there is a direct path to any negative applications, the authors should point it out. For example, it is legitimate to point out that an improvement in the quality of generative models could be used to generate Deepfakes for disinformation. On the other hand, it is not needed to point out that a generic algorithm for optimizing neural networks could enable people to train models that generate Deepfakes faster.
        \item The authors should consider possible harms that could arise when the technology is being used as intended and functioning correctly, harms that could arise when the technology is being used as intended but gives incorrect results, and harms following from (intentional or unintentional) misuse of the technology.
        \item If there are negative societal impacts, the authors could also discuss possible mitigation strategies (e.g., gated release of models, providing defenses in addition to attacks, mechanisms for monitoring misuse, mechanisms to monitor how a system learns from feedback over time, improving the efficiency and accessibility of ML).
    \end{itemize}
    
\item {\bf Safeguards}
    \item[] Question: Does the paper describe safeguards that have been put in place for responsible release of data or models that have a high risk for misuse (e.g., pre-trained language models, image generators, or scraped datasets)?
    \item[] Answer: \answerNA{} %
    \item[] Justification: The paper does not release a new model or scraped personal dataset. The benchmark is based on public, fixed software incidents and source-code artifacts.
    \item[] Guidelines:
    \begin{itemize}
        \item The answer \answerNA{} means that the paper poses no such risks.
        \item Released models that have a high risk for misuse or dual-use should be released with necessary safeguards to allow for controlled use of the model, for example by requiring that users adhere to usage guidelines or restrictions to access the model or implementing safety filters. 
        \item Datasets that have been scraped from the Internet could pose safety risks. The authors should describe how they avoided releasing unsafe images.
        \item We recognize that providing effective safeguards is challenging, and many papers do not require this, but we encourage authors to take this into account and make a best faith effort.
    \end{itemize}

\item {\bf Licenses for existing assets}
    \item[] Question: Are the creators or original owners of assets (e.g., code, data, models), used in the paper, properly credited and are the license and terms of use explicitly mentioned and properly respected?
    \item[] Answer: \answerYes{} %
    \item[] Justification: The paper cites the main prior benchmarks, datasets, harnesses, and model families used for comparison. The final artifact release includes explicit license metadata for each underlying open-source project and benchmark asset.
    \item[] Guidelines:
    \begin{itemize}
        \item The answer \answerNA{} means that the paper does not use existing assets.
        \item The authors should cite the original paper that produced the code package or dataset.
        \item The authors should state which version of the asset is used and, if possible, include a URL.
        \item The name of the license (e.g., CC-BY 4.0) should be included for each asset.
        \item For scraped data from a particular source (e.g., website), the copyright and terms of service of that source should be provided.
        \item If assets are released, the license, copyright information, and terms of use in the package should be provided. For popular datasets, \url{paperswithcode.com/datasets} has curated licenses for some datasets. Their licensing guide can help determine the license of a dataset.
        \item For existing datasets that are re-packaged, both the original license and the license of the derived asset (if it has changed) should be provided.
        \item If this information is not available online, the authors are encouraged to reach out to the asset's creators.
    \end{itemize}

\item {\bf New assets}
    \item[] Question: Are new assets introduced in the paper well documented and is the documentation provided alongside the assets?
    \item[] Answer: \answerYes{} %
    \item[] Justification: \Cref{sec:benchmark,app:benchmark-details,app:prompts}
    document the benchmark's inputs, file layout, labels, curation, and output
    contract. \Cref{app:benchmark-label-structure,app:benchmark-validity}
    provide annotation-agreement results and the scope and limitations of
    the curated data.
    \item[] Guidelines:
    \begin{itemize}
        \item The answer \answerNA{} means that the paper does not release new assets.
        \item Researchers should communicate the details of the dataset\slash code\slash model as part of their submissions via structured templates. This includes details about training, license, limitations, etc. 
        \item The paper should discuss whether and how consent was obtained from people whose asset is used.
        \item At submission time, remember to anonymize your assets (if applicable). You can either create an anonymized URL or include an anonymized zip file.
    \end{itemize}

\item {\bf Crowdsourcing and research with human subjects}
    \item[] Question: For crowdsourcing experiments and research with human subjects, does the paper include the full text of instructions given to participants and screenshots, if applicable, as well as details about compensation (if any)? 
    \item[] Answer: \answerNA{} %
    \item[] Justification: The paper does not involve crowdsourcing or recruited
    human participants. Human work consists of benchmark annotation and
    assessment of model diagnoses and patches
    (\Cref{app:benchmark-label-structure,app:evaluator-validation}).
    \item[] Guidelines:
    \begin{itemize}
        \item The answer \answerNA{} means that the paper does not involve crowdsourcing nor research with human subjects.
        \item Including this information in the supplemental material is fine, but if the main contribution of the paper involves human subjects, then as much detail as possible should be included in the main paper. 
        \item According to the NeurIPS Code of Ethics, workers involved in data collection, curation, or other labor should be paid at least the minimum wage in the country of the data collector. 
    \end{itemize}

\item {\bf Institutional review board (IRB) approvals or equivalent for research with human subjects}
    \item[] Question: Does the paper describe potential risks incurred by study participants, whether such risks were disclosed to the subjects, and whether Institutional Review Board (IRB) approvals (or an equivalent approval/review based on the requirements of your country or institution) were obtained?
    \item[] Answer: \answerNA{} %
    \item[] Justification: The study evaluates software agents and public
    software incidents; it does not collect participant data or conduct
    experiments on human subjects. The annotation and diagnosis audits assess
    software artifacts and model outputs.
    \item[] Guidelines:
    \begin{itemize}
        \item The answer \answerNA{} means that the paper does not involve crowdsourcing nor research with human subjects.
        \item Depending on the country in which research is conducted, IRB approval (or equivalent) may be required for any human subjects research. If you obtained IRB approval, you should clearly state this in the paper. 
        \item We recognize that the procedures for this may vary significantly between institutions and locations, and we expect authors to adhere to the NeurIPS Code of Ethics and the guidelines for their institution. 
        \item For initial submissions, do not include any information that would break anonymity (if applicable), such as the institution conducting the review.
    \end{itemize}

\item {\bf Declaration of LLM usage}
    \item[] Question: Does the paper describe the usage of LLMs if it is an important, original, or non-standard component of the core methods in this research? Note that if the LLM is used only for writing, editing, or formatting purposes and does \emph{not} impact the core methodology, scientific rigor, or originality of the research, declaration is not required.
    \item[] Answer: \answerYes{} %
    \item[] Justification: \Cref{sec:evaluation,app:validity-evidence} describe
    the LLM-based diagnosis evaluations, repair pilot, and closed-book probes.
    \Cref{app:collection-pipeline-eval} documents the separate LLM-assisted
    collection experiment; the released gold labels remain human-curated.
    \item[] Guidelines:
    \begin{itemize}
        \item The answer \answerNA{} means that the core method development in this research does not involve LLMs as any important, original, or non-standard components.
        \item Please refer to our LLM policy in the NeurIPS handbook for what should or should not be described.
    \end{itemize}

\end{enumerate}

\end{document}